\documentclass[a4paper,twocolumn,11pt,accepted=2017-05-09]{quantumarticle}
\pdfoutput=1
\usepackage[utf8]{inputenc}
\usepackage[english]{babel}
\usepackage[T1]{fontenc}
\usepackage{amsmath}
\usepackage{hyperref}
\usepackage{tikz}
\usepackage{lipsum}
\usepackage{mathtools}
\usepackage{amsmath}
\usepackage{amssymb}
\usepackage{graphicx}
\usepackage{physics}
\usepackage{tabularray}
\hypersetup{
    colorlinks=true,
    linkcolor=blue,
    urlcolor=blue,
    citecolor=blue,
    pdfborder={0 0 0},
}
\usepackage{bm}
\usepackage{bbold}
\definecolor{mygreen}{rgb}{0,0.5,0}
\definecolor{myblue}{rgb}{0,0,0.75}
\definecolor{mymagenta}{cmyk}{0,1,0,0.12}

\newcommand{\id}{{\sf 1 \hspace{-0.3ex} \rule[.02ex]{0.1ex}{1.52ex} \rule[.02ex]{0.3ex}{0.15ex} }}

\newcommand{\minus}{
	\setbox0=\hbox{-}
	\vcenter{
		\hrule width\wd0 height \the\fontdimen8\textfont3
	}%
}

\newcommand{\be}{\begin{equation}}\newcommand{\ee}{\end{equation}}
\newcommand{\lf}{\left(}\newcommand{\ri}{\right)}

\newcommand{\tk}{\textbf{k}}
\newcommand{\vk}{\vec{k}}

\newcommand{\hvd}{\hat{\vec{D}}}
\newcommand{\hve}{\hat{\vec{E}}}
\newcommand{\hth}{\hat{\theta}}
\newcommand{\hvp}{\hat{\varphi}}
\newcommand{\hbh}{\hat{H}}
\newcommand{\hr}{\hat{\rho}}
\newcommand{\hva}{\hat{\vec{A}}}
\newcommand{\hba}{\hat{A}}
\newcommand{\hbe}{\hat{E}}
\newcommand{\qe}{q_{\mathrm{eff}}}
\newcommand{\vn}{\vec{n}}

\definecolor{mygreen}{rgb}{0,0.5,0}\definecolor{myblue}{rgb}{0,0,0.75}\definecolor{mymagenta}{cmyk}{0,1,0,0.12}

\newcommand{\ha}{\hat{a}}

\makeatletter
\AtBeginDocument{\let\LS@rot\@undefined}
\makeatother

\begin{document}
                  	
\title{Angular Momentum Entanglement Mediated By General Relativistic Frame Dragging} 

\author{Trinidad B. Lanta\~no}\email[]{trinidad.lantano-pinto@uni-ulm.de} \affiliation{Institut f\"ur Theoretische Physik \& IQST, Albert-Einstein-Allee 11, Universit\"at Ulm, 89069 Ulm, Germany}  \author{Luciano Petruzziello}\email[]{luciano.petruzziello@uni-ulm.de}
\affiliation{Institut f\"ur Theoretische Physik \& IQST, Albert-Einstein-Allee 11, Universit\"at Ulm, 89069 Ulm, Germany} 
\author{Susana F. Huelga}\email[]{susana.huelga@uni-ulm.de} \affiliation{Institut f\"ur Theoretische Physik \& IQST, Albert-Einstein-Allee 11, Universit\"at Ulm, 89069 Ulm, Germany}
\author{Martin B. Plenio}\email[]{martin.plenio@uni-ulm.de} \affiliation{Institut f\"ur Theoretische Physik \& IQST, Albert-Einstein-Allee 11, Universit\"at Ulm, 89069 Ulm, Germany}
 
\begin{abstract}
{Current proposals to probe the quantum nature of gravity in the low-energy regime predominantly focus on the Newtonian interaction term. In this work, we present a theoretical exploration of gravitationally mediated entanglement arising from a genuinely general relativistic effect: frame dragging. This interaction gives rise to an effective dipolar coupling between the angular momenta of two rotating, spherically symmetric masses, allowing entanglement generation between angular momentum degrees of freedom. 
We represent the quantum states by angular momentum eigenstates and show that, while the maximal entangling rate is achieved for highly delocalized initial states, non-negligible quantum correlations can still emerge even when the initial states are not prepared in superposition. We then analyze the robustness of the resulting entanglement in the presence of common noise sources, explicitly acknowledging the challenges associated with a potential implementation. We also note that, for spherically symmetric masses, angular momentum degrees of freedom are intrinsically insensitive to Casimir and Coulomb interactions, thereby mitigating key decoherence channels present in existing proposals.
Finally, we discuss possible state preparation and detection strategies while framing our results within the broader landscape of gravitationally mediated entanglement schemes, emphasizing the role of this framework as a conceptual avenue for exploring genuinely relativistic quantum gravitational effects.}

\end{abstract}
	
\maketitle
	
\emph{Introduction} -- The desire to determine experimentally the character of gravity - whether 
it is quantum mechanical or classical in nature - remains an unmet challenge due to the extreme 
weakness of the gravitational interaction compared to all the other known forces. 

The earliest considerations of experimental tests date 
back at least to a thought experiment proposed by Feynman during a discussion session at the 1957 
\textit{Conference on the Role of Gravitation in Physics} at Chapel Hill, North Carolina \cite{Dewitt2011TheRO}. 
His thought experiment questioned how the gravitational field of a particle, whose center-of-mass 
degree of freedom is placed in a coherent superposition via a Stern-Gerlach apparatus, would act 
on a test mass. At the time, realizing such an experiment was entirely inconceivable, as even the control of a single quantum object such as an atom had not yet been achieved. 

Recently, this question has been revisited in the field of quantum technologies. 
Notably, in 2005, Lindner and Peres considered the gravitational field of a Bose-Einstein condensate 
in a quantum superposition and proposed probing it by scattering a particle off the condensate, which would lead to gravitationally-induced entanglement between the condensate and the particle \cite{LindnerP2005}, and thus to a low-energy test of quantum gravity.
Following these early steps, the last decade has seen an increasing number of experimental proposals 
aimed at elucidating the quantum nature of gravity in this spirit
\cite{Bose2017,Elahi2025Diamagnetic,bosermp,Marletto2017,KafriT2013,Schmoele2016,mann,Krisnanda2020,Miao2020,Chevalier2020,christodoulou2023Linearized,pedernales2020motional,yamamoto,cosco2021enhanced,weiss2021large,miki,Pedernales2022}. 

When discussing this topic and its potential tests, it is essential to agree on
how to define \textit{quantum} and \textit{classical} behavior of an interaction. To do so, we interpret the interaction between particles as giving rise to a physical channel through which information
can be exchanged and correlations may be established. A classical interaction will henceforth be 
understood as any action that can be described by local quantum operations (LO) on each of the particles and an exchange
of classical information (CC) between their locations - referred to collectively as LOCC \cite{LOCC}. A quantum mechanical 
interaction, however, allows for the coherent exchange of quantum states and can generate the most general 
quantum dynamics between the particles involved. Thus, testing the quantum nature of the object mediating 
the interaction involves determining the properties of the channel between the particles. If this channel 
can transfer an arbitrary unknown quantum state with perfect fidelity, then it is quantum; if it fails to 
exceed a certain fidelity, then it is a LOCC channel \cite{LamiPP2024}. Another test probes whether the channel
can establish entanglement between the two particles, classifying it as quantum if successful or classical if 
it only generates classical correlations. In the context of this work, a channel mediated by gravity is
considered classical if its effects can be described by LOCC, and quantum if they cannot \cite{KafriT2013,KafriTM2014,KafriMT2015}.

{Existing proposals require coherent control of masses of 
at least $10^{-15}$ Kg, prepared either in macroscopic Schr{\"o}dinger cat states \cite{Bose2017, Elahi2025Diamagnetic,bosermp} or in Gaussian wave packets with spatial delocalization of roughly 
$250$ $\mu$m \cite{Krisnanda2020, cosco2021enhanced, weiss2021large, PedernalesP23}. In addition, the particles must be separated by micrometer-scale distances to suppress short-range electromagnetic interactions, such as Casimir-Polder forces. These requirements stand in sharp contrast to current experimental capabilities: matter-wave interferometry has demonstrated center-of-mass delocalization of individual molecules with masses of around $10^{-24}$ {kg} over $\sim 260$ nm \cite{arndt2}.} 

{Moreover, most current tests of low-energy quantum gravity only probe the ability of \textit{Newtonian} gravity to convey quantum information \cite{ LindnerP2005, Marletto2017, Schmoele2016,Krisnanda2020,KafriMT2015}. Since the Newtonian potential is only a static solution of the weak-field, non-relativistic limit of Einstein's equations, the conclusions that can be drawn from such experiments are limited to \textit{that} specific regime. Importantly, they do not test dynamical or genuinely relativistic aspects of the gravitational field, and the correct interpretation of gravitationally mediated entanglement (GME) in this regime remains debated \cite{bosermp,MartinMartinezPerche2023,rovelli,Christodoulou2023,Christodoulou2024,HallReginatto2018,MarconatoMarletto2021,Pikovski2022, DiBiagio2025}. Some authors argue that, together with independent evidence that gravity behaves as a local relativistic field, detecting GME sourced by the Newton potential would already imply the quantum nature of gravity \cite{christodoulou2023Linearized}. Nevertheless, explicit proposals that go beyond the Newtonian limit remain scarce.}

{One step beyond the Newtonian regime was taken in Ref.~\cite{angularmomentum}, which considers entanglement generation via the mass-energy relation in special relativity. There, gravity correlates kinetic rotational energy eigenstates of spheres of radius $0.1$ m placed in superpositions of angular velocities of $1$ Hz. While this proposal addresses special relativistic corrections to Newtonian gravity, it still does not probe the general relativistic structure of the gravitational field itself.}

{A conceptually distinct advance from a theoretical standpoint was recently made by Chen and Giacomini \cite{giacomini}, who derived an entangling phase in linearized gravity that originates explicitly from the canonical commutation relations between gravitational and matter degrees of freedom. This contribution cannot be recovered by simply adding the Newtonian potential to a quantum Hamiltonian and therefore offers a potential route toward probing canonical quantization of the gravitational field.} 

{To further contribute to this conceptual landscape, we propose a protocol that tests entanglement mediated by a general relativistic interaction: the Lense-Thirring effect \cite{Lense}. Unlike the Newtonian regime where the gravitational source is mass, the Lense-Thirring interaction is sourced by angular momentum, thus providing a direct link between quantum mechanical rotational degrees of freedom and a genuinely general relativistic phenomenon. This opens a qualitatively new direction for probing relativistic aspects of gravity in a quantum regime.}

{An additional advantage of this framework is that position-dependent forces (such as Coulomb and Casimir-Polder interactions) do not affect entanglement generated purely between angular momentum degrees of freedom (Appendix~\ref{app:newtoncasimir}). This marks a significant distinction from spatial superposition protocols, where short-range forces represent a dominant challenge. Nonetheless, the experimental requirements of our proposal remain formidable. To address these challenges, we perform a detailed analysis of the impact of realistic noise sources on both entanglement generation and its certification.}

{Indeed, achieving the necessary conditions would require pushing the boundaries of current technology: ultralow pressures of order $10^{-17}$ Pa (achieved in antiproton experiments \cite{base}), cryogenic temperatures around $0.1$ K (similar to the requirements of Newtonian-based GME proposals \cite{Bose2017}; note that in Ref. \cite{Krisnanda2020} the required temperatures are instead of the order of K), and rotation of a levitated sphere of radius $R = 50 \mu\mathrm{m}$ at angular velocities $\omega = 10^7 \mathrm{Hz}$ (with recent demonstrations of GHz rotation in nanospheres \cite{silica}, in our case advances would be needed to achieve such fast rotations for \textit{micro}spheres). While most of these conditions have been achieved individually, their simultaneous realization in a single experiment would be unprecedented. Additionally, preparing states with well-defined angular-momentum quantum numbers as large as $l \sim 10^{23}$ is likely the most challenging requirement.}

{Finally, we note that, despite the early proposals for gravitationally mediated entanglement \cite{ LindnerP2005,Bose2017, Marletto2017, Schmoele2016,Krisnanda2020,KafriMT2015} appeared experimentally formidable, rapid progress towards ground state cooling \cite{weiss2021large,Kamba2022OpticalColdDamping,Dania2025HighPurity} and considerable efforts to produce superpositions of mesoscopic masses \cite{Bose2025SpinBased} have moved them towards the realm of feasibility. In a similar spirit, we hope that our proposal will encourage developments toward superpositions of rotational states, ultimately enabling new tests of the interplay between general relativity and quantum physics.}

      \emph{The setup} -- Consider two spherically-symmetric microspheres $A$ and $B$ that are electrically neutral, free of electric dipoles, and {diamagnetically} levitated in space. These microspheres are rotating with an angular momentum $L_A$, $L_B$ oriented along the z-direction, 
with masses $M_A$, $M_B$. The masses are held positioned along the $z$-axis at a distance $r$ from each other. The rotational energy and the gravitational interaction of the angular momenta is described  by the Hamiltonian \cite{gullu1,gullu2}
\begin{align}\label{eq:Ht}
    \hat{H} & =\, \frac{\hbar^2\hat{L}^2_{A}}{2I_A} + \frac{\hbar^2\hat{L}^2_{B}}{2I_B} - \frac{G M_A M_B}{\hat{d}}\nonumber \\ &-\frac{G \hbar^2}{c^2 \hat{d}^3} \left[\hat{\vec{L}}_A\cdot \hat{\vec{L}}_B - 3(\hat{\vec{L}}_A \cdot \vec{e_z})(\hat{\vec{L}}_B \cdot \vec{e_z}) \right], 
\end{align}
where $I_A$ and $I_B$ are the moments of inertia, $\hat{d} := {r + \abs{\hat{\vec{r}}_B- \hat{\vec{r}}_A}} $ with $r$ the distance between the spheres' centers of mass and $\hat{\vec{r}}_i$ the displacement {of mass $i$ from its} equilibrium position. The dimensionless $\vec{L}_i$ (we already factored out an $\hbar$) denotes the angular momentum of {mass $i$} 
and 
$\vec{e}_z$ is the unit vector connecting the two objects \cite{wald,gravitation}. While other relativistic corrections
exist in a fully post-Newtonian expansion of the gravitational potential \cite{weinberg}, they can be neglected when focusing on entanglement between angular momentum degrees of freedom. Note that, when considering two extended objects, a multipole expansion of the Newtonian potential is in order. {The impact of} deviations from spherical symmetry can be studied by considering the leading-order contributions (see Appendix \ref{app:newtoncasimir})
 
We introduce the { angular momentum operator for the particle $k$ as $\hat{\vec{L}}_k = \hat{L}_{kx}\vec{e_x} + \hat{L}_{ky}\vec{e_y}+ \hat{L}_{kz}\vec{e_z} $}, with $\hat{L}_{kx} = (\hat{L}_+ + \hat{L}_-)/2$, $\hat{L}_{ky} = (\hat{L}_+ - \hat{L}_-)/(2i)$. {The angular momentum eigenstate $\ket{l, m}$ satisfies}
\begin{align}
\hat{L}_z \ket{l,m} &= m  \ket{l,m}, \nonumber \\[2mm] 
\hat{L}_{\pm} \ket{l,m} &= \sqrt{l (l+1)-m (m\pm 1)} \ket{l,m \pm 1}.
\end{align}

\emph{Entanglement build-up}
-- {We assume two levitated microspheres prepared in the product state}
\begin{align}\nonumber
    \ket{\psi_{AB}} =\,&\mathcal{N}\left( \ket{l_A, m_A} + \ket{l_A, -m_A} \right)\\[2mm]\label{eq:instate}
    &\otimes\left( \ket{l_B, m_B} + \ket{l_B, -m_B} \right),
\end{align}
{where $\mathcal{N}$ is the normalization factor.} 
Since the operator $\hat{L}^2$ commutes with all its components, in the interaction picture {with respect to the kinetic part the Hamiltonian \eqref{eq:Ht} reads} 
\begin{align}\label{eq:HI}
\hat{H}_{I}=&-\frac{\alpha \hbar}{2} \left(\hat{L}_{A+} \hat{L}_{B-} + \hat{L}_{A-} \hat{L}_{B+} -4\hat{L}_{Az} \hat{L}_{Bz}\right), 
\end{align}
where $\alpha:={G \hbar}/{(c^2 r^3)}$.
For a compact notation, we henceforth write $\ket{l_i, m_i} := \ket{m}_i$
and find the state at time $t$ to be
\begin{align}\label{eq:state1}
    \ket{\psi_{AB}(t)}=\,&  \mathcal{N}\text{e}^{\frac{i \alpha t }{2} \left(\hat{L}_{A+} \hat{L}_{B-}  + \hat{L}_{A-} \hat{L}_{B+} - 4 \hat{L}_{Az} \hat{L}_{Bz}\right)} \nonumber  \\[2mm]
    &\times\left(\ket{m}_A+\ket{-m}_A\right)\left(\ket{m}_B+\ket{-m}_B\right). 
\end{align}
We compute the amount of entanglement via the Von Neumann entropy $\mathcal{S}$ \cite{Nielsen_Chuang_2010}.
Expanding the exponential operator in Eq. (\ref{eq:state1}) up to second order yields
\begin{align}\label{eq:state_ent}
    \ket{\psi_{AB}(t)} &= \Big(\mathbb{1} + \frac{i\alpha t}{2} \hat{O} _{AB} \nonumber\\ &- \frac{\alpha^2 t^2}{8} \hat{O} _{AB}  \hat{O} _{AB}\Big) \ket{\psi_{AB}(0)},
\end{align}
where $\hat{O}_{AB}:=\hat{L}_{A+} \hat{L}_{B-}  + \hat{L}_{A-} \hat{L}_{B+} - 4 \hat{L}_{Az} \hat{L}_{Bz}$. Let us define the couplings $\kappa (m, t) := {\alpha t m^2}/{2}$, where $-l \leq m \leq l$ and $g := \kappa (l, t) = {\alpha t l^2}/{2}$. By keeping only contributions up to $\kappa^2$ and $g^2$, we obtain
\begin{align} \label{eq:vn_entropy_stateAB}
    &\mathcal{S}(\rho_{A}) =  \left(2 g^2-4 g \kappa +18 \kappa ^2-1\right) \log_2 \left(-2 g^2+4 g \kappa\right. \nonumber \\[2mm] & \left. -18 \kappa ^2+1\right)-4 (g-\kappa )^2 \log_2 \left(g-\kappa\right)\nonumber\\[2mm] &-32 \kappa ^2 \log_2 \left(4 \kappa\right).
\end{align}
\begin{figure}[h!]
\hspace{-5mm}
\centering
\includegraphics[width=8.1cm]{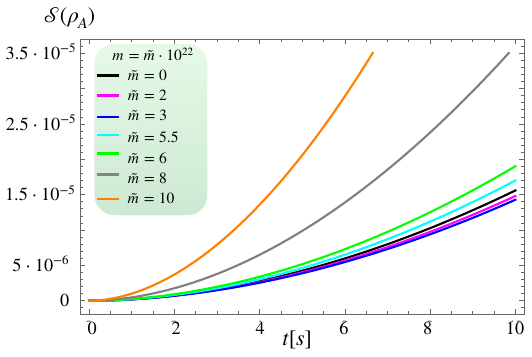}
\caption{{Time evolution of the von Neumann entropy as given by Eq. (\ref{eq:vn_entropy_stateAB}) for $l=10^{23}$. Each curve corresponds to a different $m$ quantum number as defined in Eq. (\ref{eq:state1}). While the largest entangling rate is achieved for $m=l$, even the more easily prepared $m=0$ configuration yields significant rates.}}
  \label{fig2}
\end{figure}

Higher-order contributions in $\kappa$ and $g$ are negligible for the parameters considered in this work. We consider two identical silica ($\mathrm{SiO_2}$) microspheres with mass density $\rho = 2200$ kg/m$^3$, radius $R = 50$ $\mu$m, a center-of-mass separation of $r = 4R$ and rotation velocities $\omega = 10^7$ Hz. This gives $\alpha t \approx 10^{-50} t$, where $t$ is the evolution time. The total angular momentum is $L_A = L_B = L \approx 1.15 \cdot 10^{-11}$ J$\cdot$s, yielding a quantum number $l_A = l_B = l \approx 10^{23}$. For $l \gg 1$, the ladder operators act as $\hat{L}_{+} \ket{0} \approx l \ket{1}$ and $\hat{L}_{-} \ket{0} \approx l \ket{-1}$. This leads to terms proportional to $\kappa \leq g = \alpha t l^2 / 2 \approx 6 \cdot 10^{-5} t$, ensuring the approximation holds for reasonable experiment durations.

{In choosing the above experimental parameters, we have performed an optimization to keep the entanglement growth considerable. To this aim, we note that the main parameter regulating the entanglement generation is the characteristic frequency of the analyzed system, namely $\alpha l^2$. The function $\mathcal{S}(\rho_A)$ is a monotonically increasing function of such a frequency, meaning that, to achieve a high degree of quantum correlations, one has to enhance $\alpha l^2$ as much as possible.}

{For this reason, we have provided the microspheres with a significant angular velocity, close to the largest one that can be tolerated by spinning levitated masses made of silica for reasons of mechanical stability. To show this, one can verify that the maximal allowed value for $\omega$ is related to the ultimate tensile strength of the material $\sigma_{UTS}$ as \cite{steiner}
\begin{align}\label{eq:max_speed}
    \omega_M=\sqrt{\frac{\sigma_{UTS}}{\rho R^2}}\,.
\end{align}
In the case of silica, the ultimate tensile strength is of the order of the GPa \cite{ultimatensile}, thus implying that, with our experimental parameters, $\omega_M\simeq1.6\times10^7$ Hz. Beyond this threshold, the microspheres would break due to the intense centrifugal stress.
}

{In principle, smaller masses are capable of bearing even higher values of $\omega_M$ \cite{silica}, but the extreme rotational velocity is not the only variable to take into account. Indeed, one can show that, since $\alpha\sim \mathcal{O}(1/R^3)$ and $l\sim \mathcal{O}(MR^2\omega_M )\sim \mathcal{O}(R^4)$, the characteristic frequency of the considered setup scales with the linear size of the spheres as $\alpha l^2\sim \mathcal{O}(R^5)$. Thus, even one order of magnitude less than the currently proposed one for $R$ would lead to a negligible growth rate for entanglement generation. On the other hand, an increase of the size of the spheres would also jeopardize the establishment of quantum correlations, as they would be completely suppressed by decoherence channels such as the collisional scattering.}

Concerning the initial conditions, while an initial state with $m=l$, \emph{i.e.} $(\ket{l,l}+\ket{l,-l})/\sqrt{2}$, yields the fastest entanglement growth, the eigenstate $\ket{l,0}$ still achieves appreciable entangling rates -- surpassing those for all $m \lesssim 0.95 l$. 
Importantly, preparing a microsphere in a specific angular momentum eigenstate does \emph{not} require resolving individual quanta. Instead, as shown in Appendix \ref{app:detection}, the eigenvalue $m$ of the $\hat{L}_z$ operator can be inferred from center-of-mass measurements. Specifically, the variance resolution of $\hat{L}_{kz}$ for microsphere $k \in \{A, B\}$ is given by
\begin{align}\label{var}
(\Delta L_{kz})^2(t) &=\frac{\gamma^2 I_k^2 \mathrm{G}^2_0}{4\sin^4\lf{\sqrt{\frac{|\chi_V|}{4\rho\mu_0}}\mathrm{G}_0t}\ri}\nonumber\\[2mm]
&\times[(\Delta z_k)^2(t)-(\Delta z_k)^2(0)],
\end{align}
with $\gamma\approx8$ MHz$/$T being the silica gyromagnetic ratio, $\chi_V\approx10^{-5}$ the magnetic susceptibility and G$_0$ the spatial gradient of a non-uniform magnetic field used to trap the microsphere.

{A further advantage of the probe eigenstate $\ket{l,0}$ is that its preparation is supported by recent experimental progress in levitated optomechanics. In particular, Ref.~\cite{Piotrowski2023Simultaneous2D} demonstrated that, by ground-state cooling two transverse motional modes of a levitated nanoparticle with a diameter of $\sim140 \,\mathrm{nm}$ and a mass of $\sim10^{-18}\, \mathrm{kg}$, one can prepare states with $\langle L_z \rangle \simeq 0$ and fluctuations $\sqrt{\langle L_z^2 \rangle} \approx 1.7$ $\hbar$. While this result does not imply that the final state of the system is $|l,0\rangle$ exactly, it demonstrates that it is possible to achieve a narrowly peaked distribution in the $m$ basis around the value $m=0$, which represents a first step towards the realization of the initial state for our protocol. Moreover, the same work argues that, when combined with the coherent injection of orbital angular momentum, such techniques could enable the stabilization of states with large angular momentum and quantum-limited variance.}


{Whilst the above discussion supports the possibility of engineering large angular momentum states in levitated platforms, we need a further requirement to ensure that the relevant quantum numbers remain well-defined throughout the interaction. In particular, we refer to the fact that, in the ideal spherical limit, the $2l+1$ states $\ket{l,m}$ at fixed $l$ are degenerate in $m$, thereby reflecting the absence of a preferred quantization axis. As a consequence, even faint, uncontrolled perturbations (such as residual trap anisotropies, stray fields, mechanical vibrations, etc.) may in principle induce slow diffusion or mixing among different $m$ sub-levels on relatively long timescales.} 

{Therefore, any potential experimental protocol must implement a finely-tuned ``symmetry breaking'' mechanism that selects a quantization direction and stabilizes the chosen $\ket{l,m}$ (or $\ket{l,m}\pm\ket{l,-m}$) by producing an effective splitting $\Delta E_m$ between the relevant $m$ subspaces. A minimal consistency requirement for treating $\ket{l,m}$ as a robust basis during a given evolution time $t$ is that transitions in the $m$ number are off-resonant on the experimental timescale. A simple way to convey this concept is to require
${\Delta E_m}/{\hbar} \gg {1}/{t}$.
Several physically plausible routes could provide such stabilization without altering the gravitational entanglement generation discussed so far.}

{For instance, a controlled anisotropy of the trapping potential or a deliberately introduced deviation from perfect sphericity (which however has to be subjected to the constraints analyzed in Appendix \ref{app:newtoncasimir}), can lift the $m$ degeneracy and identify a preferred axis. Alternatively, a weak external field coupling to an internal degree of freedom of the microspheres can provide an effective quantization direction, as it can already be done in laboratory tests involving nano-scale objects (see for example Refs. \cite{sta,sta2}). Although the setting of our proposal is based upon microspheres whose linear size is one to two orders of magnitude bigger than the aforementioned nano-rotors, it is not unthinkable to envisage a technological improvement capable of optimally controlling larger probes in the near future. In either case, the stabilizing factor has to be made sufficiently weak so as not to introduce significant torques while at the same time ensuring the required $\Delta E_m$ over the experimental timescale. In the present analysis, we assume that such a stabilization technique is understood and properly implemented.}

{Going back to the variance resolution \eqref{var}, one can show that, by knowing its value, it is possible to verify whether the Lense-Thirring effect is capable of establishing quantum correlations between the angular momenta of the microspheres or not. To this aim,  we rely on local sum uncertainty relations \cite{sumunc}, and in particular on} the fact that, for any separable state $\hat{\rho}_s$, the inequality 
\begin{align}\label{eq:inequality}
    \sum_\alpha \Delta (L_{A \alpha} +  L_{B \alpha})^2  \geq \hbar^2 \lf l_A + l_B \ri
\end{align}
holds, where $\alpha={x,y,z}$ and $l_k$ is the total angular momentum number of the $k$-th party. If the inequality is violated, the state must be entangled. {Therefore, one has to measure the angular momentum vector with a relative standard deviation better than $\Delta \vec{L}_{k}/L_k\approx 10^{-11.5}$, where $\vec{L}_k$ is the angular momentum vector of the $k$-th microsphere and $L_k=\hbar l_k$ its modulus. }
 Relying on the measurement scheme described in Appendix \ref{app:detection}, we can show that for any spatial direction one can achieve a resolution of the order of  inequality (\ref{eq:inequality}) when $\Delta x_\alpha\approx10^{-12}\,\mathrm{G}_0$. If 
$\mathrm{G}_0=10^4$ T$/$m {(note that Bose \emph{et al.} requires a gradient of $10^6$ T$/$m \cite{Bose2017})}, then $\Delta x_\alpha\approx10^{-8}$ m.

{This estimate indicates that the required single-shot displacement sensitivity is not the most prohibitive aspect of the proposal. The dominant limitation in the present parameter regime is the statistical certification of the entanglement witness. As a matter of fact, Eq. \eqref{eq:inequality} involves variance sums, so it cannot be established from a single test, but must rather be inferred from an ensemble of repeated runs. Under the standard assumptions of $N$ identical and independent trials, the variance scales as $\sim 1/\sqrt{N}$, thereby implying that achieving a clear violation requires very large data sets and long-term stability.} 

{Such an awareness motivates addressing the entanglement certification problem primarily from a metrological perspective. For instance, one plausible mitigation factor can be identified with optimal angular momentum state engineering. This implies preparing each rotor in suitably oriented spin-squeezed states so as to enhance the growth of quantum correlations under the action of \eqref{eq:Ht}, thus enlarging the gap to the separable bound. Another possible improvement can be given by statistical efficiency, which entails an optimization of the measurement itself with Bayesian estimation of the relevant covariance elements. This feature can in principle reduce the variance and improve the effective sensitivity of the witness evaluation.} 

{While a complete metrological analysis is beyond the scope of the present work, these considerations clarify that, once the single-shot displacement readout is within reach, the main experimental challenge shifts to the statistics of variance estimation and to protocols that maximize the detectable inequality violation within the available coherence time.}

\emph{Decoherence} -- 
{The scenario described so far assumes an idealized system, fully isolated from its environment. In our setup, using angular momentum as the degree of freedom and assuming spherical symmetry in the test masses reduces many noise sources that typically affect experiments probing the Newtonian potential via spatial delocalization or rotational degrees of freedom in asymmetric particles. The near-perfect spherical symmetry also significantly mitigates the impact of Casimir-Polder forces (Appendix \ref{app:newtoncasimir}).

We now examine three major sources of decoherence in the gravitational channel: (i) electromagnetic interaction, (ii) collisions with surrounding particles, and (iii) black-body radiation. Additionally, we explore the effects of rotating the microspheres at frequencies up to $10^7$ Hz.
Although various materials could be used for this experiment, amorphous silica microspheres -- routinely levitated and already rotated at angular frequencies up to $6$ GHz for particles with diameters of 190 nm \cite{silica} -- will serve as the basis for our experimental estimates.}

{In passing, we observe that gravitational decoherence can be safely neglected, as the typical mechanisms considered in the literature do not represent a hindrance to the entanglement growth in our protocol. As a matter of fact:
\begin{itemize}
    \item The Diosi-Penrose model \cite{dp1,dp2} is a decoherence mechanism for center of mass degrees of freedom since, in its original Newtonian formulation, it is fully characterized by the spatial mass density distribution. Therefore, rotational degrees of freedom are only affected when a change in orientation actually produces a distinguishable mass distribution in space. In the currently examined regime (\emph{i.e.}, rotational coherence of a sufficiently homogeneous and nearly isotropic rigid body), this distinction is negligible, meaning that any contribution to angular momentum decoherence is not expected to modify our analysis at leading order;
    \item Models inspired by the presence of a thermal bath of gravitons \cite{blencowe} predict a decoherence rate which is proportional to the distance in energy of the initial and the final state. However, since we are currently considering almost degenerate states, the contribution coming from this channel is negligible;
    \item Schemes of gravitational decoherence derived from linearized quantum gravity \cite{ah} and low-energy phenomenological models for quantum gravity \cite{pi,ag} have been shown to be correlated  to linear momentum dephasing, which in the current protocol is assumed to be extremely small since the spheres are trapped. Thus, decoherence of the rotational degrees of freedom appears to be of higher order according to this gravitational noise channel. 
\end{itemize}}

\noindent
Finally, we stress that the purity of the initial state is also relevant as excessive mixedness might delay entanglement build-up for too long. As long as the difference between the entropy of the reduced density matrix of one subsystem and the entropy of the global density matrix is positive, a known lower bound on the relative entropy of entanglement ensures its positivity \cite{relativent} which, in turn, implies positivity of the logarithmic negativity. 

$\bullet$  \emph{Electromagnetic interactions} -- Amorphous silica is mainly made up by neutral 
and spinless nuclei, the only source of a non-vanishing spin coming from the spin-1/2 of ${}^{29}$Si, while the 
most abundant isotope of silicon is ${}^{28}$Si. Isotope separation can reduce the ${}^{29}$Si content to 
the ppm level \cite{silicaisotope}. This would imply around $n\simeq10^9$ nuclear spins in a silica microsphere of radius $50\mu$m. 

As a result, the interaction energy associated with the magnetic dipole-dipole interaction would be 
$V_M\simeq10^{-28}p^2$ J, with $p$ the polarization. If angular momentum is of the 
order of $L\simeq10^{-11}$ J$\cdot$s, then the gravitational potential in Eq. \eqref{eq:Ht} is 
$V_G\simeq10^{-38}$ J, implying that

\begin{equation}\label{ratio}
    \frac{V_M}{V_G}\simeq p^{2} 10^{10}.    
\end{equation}
In thermal equilibrium at $T = 0.1$ K, with a magnetic field of $B = 1$ T and $\omega=10^7$ Hz, the polarization of the ${}^{29}$Si due to the Barnett effect is \cite{barpar}
$p={\hbar}\lf\gamma B+\omega\ri/\lf k_BT\ri\approx10^{-3}$,
meaning that a Faraday shield has to be inserted between the two microspheres to further suppress the electromagnetic interaction. {An optimal strategy consists in inserting a conducting screen between the two spheres. A conceptually simple implementation is made up of a thin metallic membrane placed close to the mid-plane between the microspheres, potentially supported on a frame that preserves optical access and minimizes mechanical coupling to the trap.}

{Such a screen strongly attenuates direct Coulomb interactions between the two bodies (given by, e.g., charged impurities), while leaving the gravitomagnetic channel unaffected. In addition to that, to completely shield the magnetic interaction stemming from dipole-dipole couplings, the Faraday shield has to be further coated with a superconducting material. Such a procedure has already been engineered in several experiments \cite{shield1,shield2}: The screening is achieved through gold-plated copper shields which have been enveloped with a superconducting lead foil and a final Cryoperm layer to deal with the extremely low temperature required by the laboratory test.}

{We observe that, in principle, a realistic shield introduces additional constraints, such as Casimir forces between each sphere and the membrane as well as weak gravitational and vibrational backgrounds associated with the screen \cite{BullingSteinerPedernalesPlenioInpreparation}. However, since the witness in Eq. \eqref{eq:inequality} is constructed from angular momentum variances and shield-induced forces act predominantly on the center-of-mass coordinate, we expect that the entanglement generation is not significantly affected, although this interaction might still indirectly reduce the available coherence time by jeopardizing the stabilization of the quantization axis.
}

{As a further issue related to the electromagnetic interactions, we stress that,} even though $\mathrm{SiO_2}$ microspheres can be prepared to be charge neutral, they can have {an electric} dipole moment due to an inhomogeneous charge distribution. Currently, the typical permanent dipole moment for such spheres is $\sim$ 100 e $\mu$m \cite{afek}, resulting in a dipole-dipole interaction energy of $V_{\text{dip-dip}}\simeq 2.8 \cdot 10^{-25}$ J, which significantly exceeds $V_G$.
To address this, we propose rotating the sphere around an axis orthogonal to the dipole's orientation, which must be previously measured \cite{afek}. At high rotation speeds, this setup ensures that all components of the time-averaged dipole moment are significantly reduced. A detailed estimation (see Appendix \ref{app:dipole}) yields that a rotational velocity  
of $10^7$ Hz reduces the potential energy to $V_{\text{dip-dip}}\simeq  10^{-40}$ J $< V_G$. Furthermore, our analysis shows that allowing for an angular deviation of $\Delta \delta \simeq 10^{-7}$ in the dipole orientation still ensures $V_{\text{dip-dip}} \simeq  10^{-39}$ J, which remains below $V_G$.

$\bullet$ \emph{Collisions} -- 
In the following, we focus on a single collision and require that such an event can change only the $m$ number by an amount $q$ up to a maximum $n<l$ while leaving the $l$ number untouched. If the probability of $k$ of such events happening during a time interval $t$ is described by the Poisson distribution $P(k;t)={(r t)^k \text{e}^{-r t}}/{k!}$, where $r$ is the average rate of events, the state will be given by (up to normalization)
\begin{align}\label{eq:mixed1} 
\rho_{AB} =  P\left(0;t\right) \rho_0  + P \left(1;t\right) \sum_{q=1}^n \sum_{k=\pm} \lf \rho_{Ak}^q + \rho_{Bk}^q \, \ri,
\end{align}
where
\begin{align}
    \rho_0(t) &= \ketbra{\psi_{AB}(t)}{\psi_{AB}(t)}, \nonumber \\[2mm]   \rho_{A\pm}^q(t) &= (\hat{L}_{A\pm})^q \ketbra{\psi_{AB}(t)}{\psi_{AB}(t)} (\hat{L}_{A\mp})^q ,
\end{align}
with $\ket{\psi_{AB}(t)}$ given by Eq. (\ref{eq:state_ent}) and where we have assumed $q$ to be uniformly distributed.
To obtain a rough estimate of the collision rate $r$ between $H_2$ molecules and the microsphere, we use the kinetic theory of gases and the ideal gas law to write $r=\pi R^2 P/\sqrt{2 \pi k_B T m}$, where $k_B$ the Boltzmann constant, $R$ the radius of the sphere, $T$ and $P$ the gas temperature and  pressure respectively, and $m$ the molecular mass of $\text{H}_2$. For a microsphere of radius $R=50 \mu$m and a gas temperature and pressure of $T=0.1$ K and $P=10^{-17}$ Pa respectively, the collision rate with $H_2$ molecules is $r\approx 0.4$ s$^{-1}$. Note that pressures below $10^{-16}$ Pa are achieved in antiproton spectroscopy experiments \cite{Latacz2024, Sellner2017}.

\begin{figure}[h!]
\hspace{-1mm}
\centering
\includegraphics[width=6.5cm]{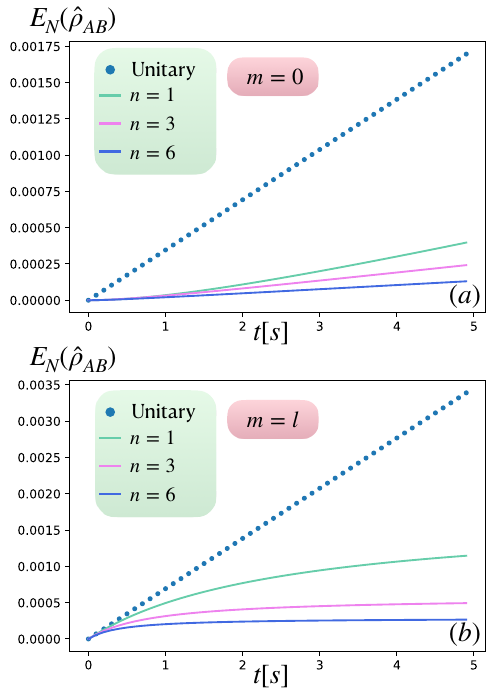}
\caption{Logarithmic negativity of state (\ref{eq:mixed1}) as a function of time considering the total number of exchanged quanta in a single collision to be $n=\{1,3,6\}$. We study the decoherence of the state given by Eq. (\ref{eq:state1}) in two cases: (a) $m = 0$ and (b) $m = l$.}
  \label{fig3}
\end{figure}

We quantify entanglement using the Logarithmic Negativity $E_N(\rho_{AB}(t))$ \cite{negativity} and present numerical results for two initial states: one with $m=0$ and the other with $m=l$ in Eq. (\ref{eq:instate}). In both cases, the total angular momentum is set to $l=10^{23}$ (see Fig. \ref{fig3}).
We observe that even a single collision, with no change in the $l$ number, causes a sharp drop in entanglement. Therefore, to preserve entanglement, the probability of a collision must be kept well below unity.

$\bullet$ \emph{Black-body radiation} -- Apart from collisions with particles from the background gas, also the 
emission or absorption of a photon has the potential to change the angular momentum quantum numbers $l$ and $m$. Hence,
the impact of black-body radiation on the evolution of Eq. (\ref{eq:state1}) needs to be analyzed carefully.

We describe the absorption and emission of thermal photons in an open quantum system approach. While this method aligns conceptually with \cite{stickler,stickler3,stickler2,stickler4}, it is important to note that our procedure focuses on angular momentum degrees of freedom rather than orientational ones (\emph{i.e.}, Euler angles).

Regarding the system
density matrix $\hr_S$ defined on the Hilbert space $\mathcal{H}_A\otimes\mathcal{H}_B$, see Appendix \ref{app:blackbody} for a detailed 
derivation of a quantum master equation which describes the joint evolution of spheres $A$ and $B$ as they undergo the 
gravitational interaction (\ref{eq:Ht}) along with a dissipative evolution caused by a thermal bath. Assuming that the characteristic wavelength of
the photons is larger than the separation between the masses, the equation reads
\begin{align} \label{eq:ME}
    \frac{d}{dt}\hr_S(t) =&-\frac{i}{\hbar}\left[\hbh_I^{AB},\hr_S(t)\right]\\ \nonumber
    &\hspace{-0.5cm}+ \sum_{l\geq 0}\sum_{p} \left[\hat{C}_{l,p} \hr_S(t) \hat{C}_{l,p}^\dagger - \frac{1}{2} \{ \hat{C}_{l,p}^\dagger \hat{C}_{l,p},\hr_S(t) \}\right]
    \\ &\hspace{-0.5cm}+  \sum_{l\geq 0}\sum_{p} \left[\hat{F}_{l,p} \hr_S(t) \hat{F}_{l,p}^\dagger - \frac{1}{2} \{ \hat{F}_{l,p}^\dagger \hat{F}_{l,p},\hr_S(t)\} \right],
    \nonumber
\end{align}
where $p=\{1,2,3\}$, $\{\cdot ,\cdot \}$ denotes the anticommutator,  $\hat{C}_{l,p} :=\sqrt{\chi_l}\hba^{AB}_p (\Delta_l)$ 
and $\hat{F}_{l,p} :=\sqrt{\gamma_l}\hba_p^{AB\dagger} (\Delta_l)$ are collapse operators, with
$\hva^{AB} = \sum_i \hat{A}_i \vec{v}_i \otimes \id + \id \otimes  \sum_i \hat{A}_i \vec{v}_i$ and $\hat{A}_i$ given in Eqs. (37-39) of a. The rates are 
\begin{align}\nonumber \label{eq:rates}
\chi_l &:= \frac{\Delta_l ^3 \hbar^2}{6 c^3 I^3 \epsilon_0}\lf1+N(\Delta_l)\ri d_{\text{eff}}^2\,,  \\
\gamma_l &:= \frac{\Delta_l ^3 \hbar^2}{6 c^3 I^3 \epsilon_0}N(\Delta_l) d_{\text{eff}}^2\,.
\end{align}
Here, $\Delta_l := 2 (l+1)$ with the moment of inertia $I$ of one microsphere, 
$N(\cdot)$ the bosonic mean occupation number and $\epsilon_0$ the
permittivity of free space.

To determine $\gamma_l$ and $\chi_l$ we need the effective dipole moment $d_{\text{eff}}$. 
As the amorphous silica microspheres are dielectric, we assume a
polarization $P = \epsilon_0 (\epsilon_r - 1) E(T, \lambda)$, where $\epsilon_r$ is the relative permittivity of the material and $E(T, \lambda)$ 
is the electric field amplitude of the thermal radiation at temperature 
$T$ and wavelength $\lambda$. On the other hand, the polarization can be seen as the density of dipoles in a certain volume $V$, that is, $P = d_{\text{eff}}/V$.
Therefore, an estimation of the effective dipole moment of each sphere is given by 
$d_{\text{eff}} = V (\epsilon_r - 1) \sqrt{2\epsilon_0 u(T,\lambda)}$, where we have introduced the energy density
$u(T,\lambda)=E^2(T,\lambda) \epsilon_0/2$. Using Planck's law for $u(T,\lambda)$ and Wien's law to replace the wavelength by the one that gives the maximum energy density $\lambda_{peak}=b/T$ where $b\approx 2.8\cdot 10^{-3}$ m $\cdot$ K, we have
\begin{align}
d_{\text{eff}} =  \sqrt{\frac{32\pi^2 V^2 (\epsilon_r - 1)^2 c \hbar \epsilon_0 T^5}{b^5 \lf \text{e}^{\frac{2 \pi \hbar c}{b k_B}} - 1\ri} }.
\end{align}
\begin{figure}[t!]
\hspace{-5mm}
\centering
\includegraphics[width=6.5cm]{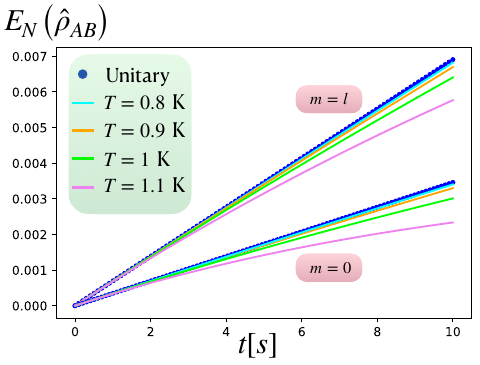}
\caption{Logarithmic negativity of the state given by Eq. (\ref{eq:ME}) as a function of time. We study the black-body decoherence of state (\ref{eq:state1}) in two cases: (lower) $m = 0$ and (upper) $m = l$.}
  \label{fig4}
\end{figure}

With the knowledge of $d_{\text{eff}}$ and the physical details of the microspheres introduced before, we have that, when the state is prepared with $m=l$ in Eq. (\ref{eq:instate}) and evolved according to Eq. (\ref{eq:ME}) at $T=0.6$ K, the decoherence rate is $\gamma_l \approx 2\cdot 10^{-4}$ s$^{-1}$, while the variation of logarithmic negativity under unitary evolution after $t=10$ s is $\Dot{E}_N := \Delta E_N(\hr_S)/\Delta t \approx 7 \cdot 10^{-4}$ s$^{-1}$.  This indicates that, at $T=0.6 $ K, the black-body radiation decoherence will not dominate over entanglement generation. However, a small increase in temperature, e.g., T=$0.8$ K, already gives $\gamma_l \approx 0.001$ s$^{-1}$, which surpasses the entanglement rate. Remarkably, the fact that $\gamma_l$ surpasses $\Dot{E}_N$ does not result in a complete suppression of entanglement build-up. Indeed, Fig. (\ref{fig4}) shows that, for T=$0.8$ K, the curve deviates slightly from the unitary evolution, and even for $T=1.1$ K, where $\gamma_l\approx 0.008$ s$^{-1}$ -- one order of magnitude larger than $\Dot{E}_N$ -- we do not observe a significant drop in the entanglement rate. 

Given that we assume both particles interact with the same bath, one might wonder whether any entanglement arises due to the radiation field. To explore this possibility, we simulated Eq. (\ref{eq:ME}) with $\hat{H}_I^{AB}=0$ and observed that the negativity remained zero over the time scales considered.

\begin{figure}[t!]
\hspace{-5mm}
\centering
\includegraphics[width=7cm]{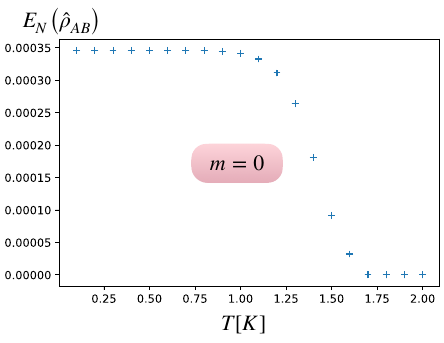}
\caption{Logarithmic negativity of an initial state $\ket{l,m=0}$ evolved in time via Eq. (\ref{eq:ME}) as a function of temperature at a fixed time $t=1$ s.}
  \label{logneg_temp1}
\end{figure}

An important question is how much entropy the system can tolerate before entanglement is completely lost. Figure \ref{logneg_temp1} shows the entanglement -- quantified by the logarithmic negativity $E_N$ -- as a function of the bath temperature, for the state $\ket{l, m=0}$ evolving under Eq.~\eqref{eq:ME} for a duration of $t = 1$ s. We observe that entanglement begins to decrease noticeably around $T = 1$ K, and becomes effectively negligible by $T \approx 1.7$ K.  At $T = 1$ K, the entropy of the global state is numerically found to be $\mathcal{S}(\rho_{AB}) \approx 0.04$ bits. When entanglement is completely lost, at $T \approx 1.7$ K, the entropy reaches approximately $\mathcal{S}(\rho_{AB}) \approx 0.6$ bits. \\
This shows that even one bit of entropy can be enough to suppress entanglement entirely. In other words, the system must be prepared in an almost pure state for entanglement to survive. While this sets a stringent requirement, recognizing this fragility is essential for designing protocols to make the state more resilient against noise. One possible strategy is to increase the effective energy gap between relevant levels, for instance by using smaller particles (and increase the angular velocity accordingly), or coupling to external fields that lift degeneracies and energetically penalize unwanted excitations. Enhancing the energy scale of the relevant degrees of freedom would reduce thermal population of higher levels and help maintain coherence. Although challenging, identifying the entropy threshold for entanglement loss provides a concrete benchmark to guide improvements in state preparation and environmental control.

$\bullet$ \emph{Laser heating} -- 
Finally, it is important to note that high-frequency rotations require the use of lasers \cite{silica}, which in turn cause heating of the spheres. This can lead to a rapid increase in temperature, reaching undesirable levels. {This is also the reason why we prefer diamagnetic over optical levitation, as in the former case the control over the temperature is more optimal than in the latter scenario. Furthermore, to keep the temperature as low as possible, one can rely on the use of a dilution refrigerator \cite{diluref}, which is capable of cooling probe systems up to the order of mK.} 

To estimate the increase in $T$, one can use the following equation \cite{steiner,bateman}:
\begin{equation}
\int_{T_i}^{T_f}dTc_M(T)=\frac{4\omega_f\lambda^2}{a\,R\rho}\mathrm{Im}\left[\frac{\varepsilon-1}{\varepsilon+2}\right],
\end{equation}
where $T_i$ ($T_f$) is the initial (final) temperature before (after) the application of the laser of wavelength $\lambda$ which brings the frequency rotation of a sphere with radius $R$, density $\rho$, specific heat capacity $c_M(T)$ and refractive index $n$ (such that $\varepsilon=n^2$) up to $\omega_f$, with $a=8.15\cdot10^{-11}$ m$^4$/(W$\cdot$s$^2$). Since we are considering low temperatures, the Debye model \cite{debye} is a good approximation for the behavior of $c_M$ as a function of $T$, that is, $c_M(T)=\beta T^3$, with $\beta\approx3\cdot10^{-4}$ J/(Kg$\cdot$K$^4$) for the amorphous silica \cite{iler}. With an initial temperature of $1$ K, a desired {rotation} velocity of $10^7$ Hz, a range of wavelength for the laser slightly outside the visible spectrum (\emph{i.e.}, $\lambda=300$ nm) and a refractive index of $n=1.47+i(0.01\lambda/(4\pi))$ \cite{steiner,iler}, it is straightforward to prove that $T_f\approx1.13$ K, thus implying that, according to the previous analysis of radiative decoherence, the disturbance due to the use of the laser is not substantial.
\begin{figure}[h!]
\centering
\includegraphics[width=9cm]{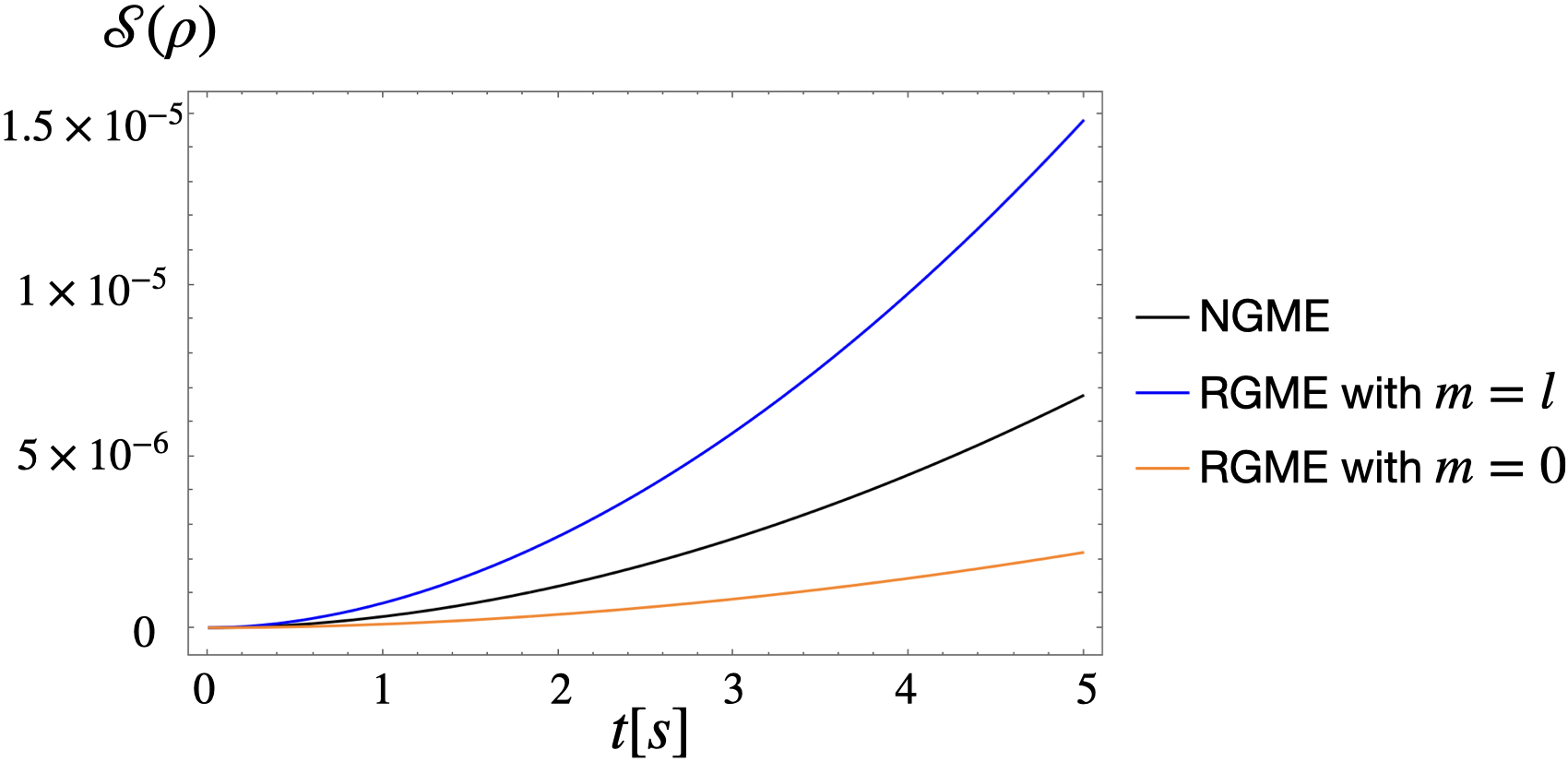}
\caption{Von Neumann entropies generated using the parameters in Table $1$. For Newtonian Gravitationally Mediated Entanglement (NGME) we selected Bose et. al \cite{Bose2017} and our work is under Relativistic Gravitationally Mediated Entanglement (RGME). For the mass in the case of NGME, we have chosen the value $10^{-15}$ kg, also used in \cite{Elahi2025Diamagnetic}.}
  \label{logneg_temp}
\end{figure}

{ \emph{Comparison with existing schemes and discussion} -- 
In order to place our work in context with the relevant literature on GME, we compare our scheme with the ones of Bose \emph{et al.} \cite{Bose2017}, Krisnanda \emph{et al.} \cite{Krisnanda2020} and Higgins \emph{et al.} \cite{angularmomentum} (see Table \ref{tab:comparison-NGME-RGME}).}
\begin{table*}[t]
\centering
\small
\begin{tblr}{
    width = \textwidth,
    colspec = {|l|X[c]|X[c]|X[c]|X[c]|},
    row{1} = {font=\bfseries},}
    \hline
    Experimental parameter 
      & Bose et al.\ 2017 
      & Krisnanda et al.\ 2020 
      & Higgins et al.\ 2024 
      & This work \\
    \hline
    Mass 
      & $ 10^{-14} - 10^{-15}\,\mathrm{kg}$ 
      & $10^{-7}\,\mathrm{kg}$
      & $\sim 96 \,\mathrm{kg}$
      &$ \sim 10^{-9}\,\mathrm{kg}$ \\
    \hline
    Radius 
      & $\sim 1\,\mu\mathrm{m}$ 
      & $ 100\, \mu \mathrm{m}$  
      & $ 0.1\,\mathrm{m}$ 
      & $50\,\mu\mathrm{m}$ \\
    \hline
    Distance between centers of mass 
      & $ 450\,\mu\mathrm{m}$
      & $ 300\, \mu \mathrm{m}$ 
      & $\sim 0.2\,\mathrm{m}$ 
      & $200\,\mu\mathrm{m}$ \\
    \hline
    Angular velocity 
      & not relevant 
      & not relevant
      & $\omega_{\max} \simeq 2\pi\,\mathrm{Hz}$ 
      & $10^{7}\,\mathrm{Hz}$ \\
    \hline
    Initial superposition 
      & macroscopic superposition: $\Delta x \simeq 250\,\mu\mathrm{m}$ 
      & Gaussian spread: $\Delta x \simeq 10^{-16} \mathrm{m}$
      & superposition of rotational orientations
      & $m$-number superposition, not necessary \\
    \hline
    Temperature 
      &  $\sim 0.1\,\mathrm{K}$ 
      & $2-4\,\mathrm{K}$ 
      & not specified; requires ultra-low $T$ 
      & $0.1\,\mathrm{K}$ \\
    \hline
    Pressure 
      & $10^{-15}\,\mathrm{Pa}$ 
      & $\sim 10^{-15}\,\mathrm{Pa}$ 
      & not given numerically; ultra-high vacuum required 
      & $10^{-17}\,\mathrm{Pa}$ \\
    \hline
\end{tblr}
\caption{Comparison of suggested parameter regimes for selected proposals of gravitationally mediated entanglement based on Newtonian gravity \cite{Bose2017,Krisnanda2020} and relativistic gravity \cite{angularmomentum}.}
\label{tab:comparison-NGME-RGME}
\end{table*}

{In Fig.~\ref{logneg_temp}, we compare the entanglement generation (quantified by the von Neumann entropy) of our relativistic gravitationally mediated entanglement (RGME) scheme with that of Ref.~\cite{Bose2017}, taken as a representative realization of Newtonian GME (NGME). Remarkably, although the interaction in the relativistic case arises only as a post-Newtonian correction, we find that comparable levels of entanglement can nevertheless be achieved under the same idealized conditions (and assuming a mass of $10^{-15}$ Kg in the NGME scenario). This is enabled by the possibility of reaching extremely large angular momenta, which provide a strong enhancement of the gravitomagnetic coupling. Importantly, the angular velocity chosen in our analysis ($\omega = 10^{7}$ Hz) is not arbitrary: as shown in Eq.~\eqref{eq:max_speed}, it lies within the experimentally achievable range that silica microspheres of the required dimensions can sustain without structural damage.}

{Another relevant aspect to highlight is that the magnitude of the generated entanglement roughly remains the same even when the initial state does not involve a superposition of rotational eigenstates. For instance, for the $m=0$ state in Eq.~(\ref{eq:instate}) corresponding to the product state $\ket{l_A,0}\otimes\ket{l_B,0}$, we see in Fig. \ref{logneg_temp} that the entanglement entropy remains within the same order as in NGME-based tests.}

{Yet, we acknowledge that preparing microspheres in well-defined rotational eigenstates constitutes a formidable experimental challenge. However, we argue that the limited experimental progress in this direction is due in part to the absence of a compelling theoretical motivation. By demonstrating that such states allow for a qualitatively new test of gravity, one that directly probes a genuinely general relativistic contribution as a quantum channel, we aim to provide this motivation. In this sense, our work does not seek to compete with existing NGME proposals in terms of near-term feasibility, but rather to establish a conceptually distinct and complementary pathway for investigating the quantum nature of gravity.}

{\emph{Roadmap towards experimental realization} -- Despite the challenges explored in detail in the manuscript, our goal is to clarify why this scheme represents a meaningful and timely contribution to the exploration of low-energy quantum gravity.
Firstly, our approach directly targets a genuine general relativistic phenomenon (frame dragging) in a quantum setting. In contrast to previous proposals that rely on Newtonian or special relativistic gravity, this scheme probes whether inherently relativistic corrections appear in quantum mechanics and can leave detectable imprints on entanglement. This provides a unique experimental pathway to testing the interplay between the two most fundamental theories of modern physics. Secondly, the protocol constitutes a novel route to gravitationally generated entanglement between distinct degrees of freedom. As we have demonstrated, this reduces the impact of decoherence channels such as Casimir interactions (albeit introducing different issues). The conceptual value of exploring such alternative gravitational couplings is independent of the technological effort required.}

{ We realize that detecting entanglement mediated by frame dragging (a subtle relativistic effect only recently observed in satellite missions \cite{framedr}) may appear infeasible at first sight. However, our analysis proves that it becomes experimentally relevant within a clearly defined parameter regime. To provide a better intuition along this direction, consider the following estimation: The product $\omega R$ for our rotating microspheres (with $\omega R \simeq 500$ m/s) is comparable to the Earth's $\omega_\oplus R_\oplus \simeq 460$ m/s. This illustrates that the gravitomagnetic field generated in our setup is not negligible when compared with what seems a more reasonable setup to observe frame dragging.}
{The above observation is crucial, as modern table-top platforms offer a level of precision and control far exceeding that of satellite experiments. When combined with the ability to reach GHz rotational frequencies in silica nanospheres, and with new technological progress in taking this to the quantum regime, this opens a window in which frame-dragging induced entanglement could be witnessed.}

{A staged experimental program toward the regime discussed here might proceed as follows. A first milestone would be to achieve and stabilize rotation frequencies approaching $10^7$ Hz for microspheres in the $10$ $\mu$m range, extending current records obtained with smaller particles. A second, classically verifiable step would involve detecting the frame-dragging torque exerted by one rapidly rotating sphere on a nearby test body, a tabletop analogue of the Gravity Probe B \cite{ciufolini} measurement that would confirm the gravitomagnetic coupling without requiring quantum state preparation. Subsequent stages would address the preparation of well-defined angular momentum eigenstates, characterization of rotational decoherence times and implementation of the magnetic-gradient readout scheme. Only once these intermediate benchmarks are met would a full entanglement test become viable.}

\emph{Final remarks} -- 
In this Letter, we introduce a new protocol for testing gravitationally-induced entanglement based on the post-Newtonian interaction \eqref{eq:Ht}, which entangles angular momentum rather than positional degrees of freedom. We demonstrate that, under suitable conditions, entanglement generation remains significant despite decoherence, even when the initial states are not in quantum superposition.

It is worth noting that using angular momentum for gravitational entanglement generation has also been proposed in Ref. \cite{angularmomentum}, where superpositions of high rotational energies and the mass-energy equivalence principle are used to enhance entanglement growth through the Newtonian potential. While the latter scheme tests a special relativistic aspect of gravity, our proposal is -- to our knowledge -- the first in testing \textit{general relativistic} effects in quantum mechanics.

{{\bf Acknowledgements} -- This work was supported by the ERC Synergy Grant HyperQ (Grant No. 856432), the EU project QuMicro (grant no. 01046911) and the DFG via QuantERA project LemaQume (Grant No. 500314265). We acknowledge discussions with Julen S. Pedernales, Benjamin A. Stickler and M. O. E. Steiner. {We are grateful to the anonymous reviewers for the insightful comments which improved the quality of the manuscript.}

\appendix
              
\begin{widetext}

 \section*{Appendix}

\section{Effect of Newton and Casimir potentials on angular momentum entanglement} \label{app:newtoncasimir}
\noindent A quantization of the gravitational potential implies that both the angular momentum and the position of the spheres are operators in Eq. (1) of the main text. 
A complete picture must contain the position states of the sphere as well. Therefore, let us assume an initial state of the spheres $\ket{\psi(0)} = \ket{\vec{r}_A}\otimes\ket{\vec{r}_B}\otimes\ket{\phi(l_a)}\otimes\ket{\phi(l_b)}$, which undergoes the action of gravity via
\begin{align}
    \hat{U} =\, \exp \left( \frac{it}{\hbar}\left[\frac{\hbar^2\hat{L}^2_{A}}{2I_A} + \frac{\hbar^2\hat{L}^2_{B}}{2I_B} - \frac{G M_A M_B}{\hat{d}} -\frac{G \hbar^2}{c^2 \hat{d}^3} \left[\hat{{L}}_{Ax}\hat{{L}}_{Bx} + \hat{{L}}_{Ay}\hat{{L}}_{By} - 2 \hat{{L}}_{Az}\hat{{L}}_{Bz} \right] \right] \right), 
\end{align}
where we placed the spheres along the $z$ - axis and we defined $\hat{d} := \abs{r+\hat{r}_B- \hat{r}_A} = r + \hat{r}_B- \hat{r}_A$, with $r$ the distance between the center of mass of the spheres and $\hat{\vec{r}}_i$ the displacement from their equilibrium position. Throughout all our computations, we assume small deviations from the center of mass position, \emph{i.e.}, $\abs{\hat{r}_i}\ll r$. Thus, we can expand the Newtonian term as follows
\begin{align}
\frac{1}{ r+ \hat{r}_B- \hat{r}_A } \approx \frac{1}{r} - \frac{ \hat{r}_B - \hat{r}_A} {r^2} + \frac{(\hat{r}_B - \hat{r}_A)^2}{r^3},
\end{align}
and analogously the relativistic term as 
\begin{align}
\frac{1}{\hat{d}^3}  \approx \frac{1}{r^3}.
\end{align}
Then, one has
\begin{align}\label{eq:evol_op}\hspace{-12mm}
    \hat{U} \approx \, \exp \left( \frac{it}{\hbar}\left[\frac{\hbar^2\hat{L}^2_{A}}{2I_A} + \frac{\hbar^2\hat{L}^2_{B}}{2I_B} + \frac{G M_A M_B}{r^2} \left(\hat{r}_B - \hat{r}_A \right) - \frac{G M_A M_B}{r^3} \left(\hat{r}_B - \hat{r}_A \right)^2 -\frac{G \hbar^2}{c^2 r^3} \left[\hat{{L}}_{Ax}\hat{{L}}_{Bx} + \hat{{L}}_{Ay}\hat{{L}}_{By} - 2 \hat{{L}}_{Az}\hat{{L}}_{Bz} \right] \right] \right). 
\end{align}
Since we are considering displacements in three dimensions, we have $\hat{r}_i = \sqrt{\hat{x}_i^2 + \hat{y}_i^2 + \hat{z}_i^2}$ with $i=\{A,B\}$. 
 By using the commutation relation
$\comm{\hat{x}_l}{\hat{L}_s}= i \epsilon_{sjl} \hat{x}_j$ where $\hat{x}_l$, $\hat{L}_s$ are the displacement and angular momentum components in direction $l$ and $s$ respectively, one can show that 
\begin{align}
    &\comm{\hat{r}_i^2 }{L_{is}}  = 0, \nonumber\\[2mm] 
    &\comm{\hat{r}_i^2 }{L_{i}^2} = 0,
\end{align}
for $s=\{x,y,z\}$, and  $i=\{A,B\}$. This result allows us to separate exactly the evolution operator into two parts: one acting on position states and the other on angular momentum states. This shows that no entanglement can arise between position and angular momentum, and therefore a potential energy containing only position operators will not be detrimental for entanglement of angular momentum degrees of freedom.

This argument can be easily extended to the Casimir forces. For two spheres of equal radius $R$, dielectric constant $\epsilon$ and center of mass separation $r$, the Hamiltonian of the Casimir-Polder potential is given by \cite{bose}
\begin{align}
   \hat{H}_C = -\frac{23 \hbar c R^6}{4\pi ( r + \hat{r}_B- \hat{r}_A)^7} \left(\frac{\epsilon-1}{\epsilon + 2} \right)^2.
\end{align}
As we only have position operators, and they commute with angular momentum in every direction, it follows that the action of the Casimir-Polder Hamiltonian will not affect the entanglement of angular momentum states.

\subsection*{Deviations from spherical symmetry}

\noindent
\begin{figure}[h!]
\hspace{-1mm}
\centering
\includegraphics[width=5.5cm]{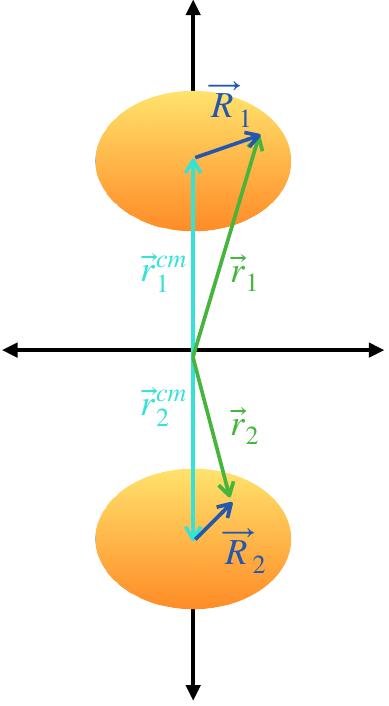}
\caption{Two oblate spheroids interact via Newtonian gravity.}
  \label{fig:Spheroid}
\end{figure}
Let us begin with the gravitational energy between two extended bodies $A$ and $B$.  The mass density of bodies $i=\{A, B\}$ is $\rho_i(\vec{R}_i)$, where $\vec{R}_i$ is the vector from the center of mass to a mass element of the corresponding body. If $\vec{r}_i$ is the vector from the origin {of the coordinate system} to a mass element of $i$, and $\vec{r}_i^{cm}$ from the origin {of the coordinate system} to the center of mass of the corresponding body (See Fig. (\ref{fig:Spheroid})), we {find the gravitational interaction energy} 
\begin{align}
    V_G^{AB}=-G\int_A \int_B \text{d}^3 \vec{R}_A  \text{d}^3 \vec{R}_B \frac{\rho_A\left(\vec{R}_A \right)\rho_B\left(\vec{R}_B \right)}{\abs{\vec{r}_A - \vec{r}_B}}.  
\end{align}
Noting that $\vec{r}_i=\vec{r}_i^{cm} +\vec{R}_i$ and defining $\vec{r}=\vec{r}_A^{cm} - \vec{r}_B^{cm}$, upon assuming a uniform density for the two bodies we have
\begin{align}
    V_G^{AB}=-G\rho_A \rho_B\int_A \int_B \text{d}^3 \vec{R}_A  \text{d}^3 \vec{R}_B \frac{1}{\abs{ \vec{r} + \vec{R}_A - \vec{R}_B}}.  
\end{align}
We expand $\abs{ \vec{r} + \vec{R}_A - \vec{R}_B}^{-1}$ for $r \equiv \abs{\vec{r}}\gg \abs{\vec{R_A}}, \abs{\vec{R_B}}$ up to $O\left( {1}/{r^3}\right)$, since these are the lowest-order terms for which deviations of spherical symmetry contribute to the potential. Defining $V_G^{AB} \approx V_G^{(1)} + V_G^{(2)} +  V_G^{(3)} + O(1/r^4)$, we have 
\begin{align} 
    V_G^{(3)} = -\frac{G\rho_A \rho_B}{2r^3}\int_A \int_B \text{d}^3 \vec{R}_A  \text{d}^3 \vec{R}_B \left[ 3\left(R_A \cos{\theta_A}-R_B\cos{\theta_B} \right)^2 - \abs{\vec{R}_A - \vec{R}_B}^2 \right],
\end{align}
where $R_i \equiv \abs{\vec{R}_i}$ and $\theta_i$ is the angle between $\vec{R}_i$ and the line that connects both center of masses. Note that we have not considered $V_G^{(1)}$ and $V_G^{(2)}$ as the first is the monopole contribution so there is no notion of size, and the second vanishes for a spheroid.\\
For a spheroid with mean radius $a$, the outer boundary satisfies
\begin{align}\label{app:radiustheta}
   R_\theta (\theta)=a \left[1 - \frac{2}{3}\varepsilon P_2 \left( \cos{\theta} \right) \right],
\end{align}
where $\theta$ is the polar angle, $P_2$ the second-order Legendre polynomial and $\varepsilon$ the ellipticity. Carrying out the integrals in spherical coordinates using Eq. (\ref{app:radiustheta}) as an upper limit for the radial integral and considering the same parameters for spheroids $A$ and $B$, the magnitude of   $V_G^{(3)}$ to leading order in $\epsilon$ is
\begin{align}
 \abs{V_G^{(3)}} = \frac{G M_A M_B a_A a_B \varepsilon_A \varepsilon_B}{512 r^3} \simeq{5.4 \cdot 10^{-29} \varepsilon^2} \text{J},
\end{align}
where $M_A=M_B={4 \pi\rho_s R^3}/{3}$ are the masses of the spheroids with radius $R=5 \cdot 10^{-5}$ m, density $\rho_s = 2200 $ kg/m$^3$, ellipticities $\varepsilon_A=\varepsilon_B=\varepsilon$ and assume a mean radius $a_A =a_B=R$ for estimation purposes.\\ 
As mentioned in the main text, the gravitational energy due to angular momentum has a magnitude of \(\abs{V_G} \simeq 10^{-38}\) J. This implies that the ellipticity must be below \(\varepsilon \simeq 10^{-5}\) for the interaction energy due to the ellipticity to remain subdominant. In other words, the difference between the two axes of the spheroid must be on the order of \(5 \times 10^{-10}\) m, approximately the size of an atom.\\
As a last remark, note that we have studied the effects of a spheroidal shape for a \textit{static} body. However, our analysis represents a worst-case scenario, since when the spheres rotate at rates much higher than the interaction rate, the relevant quantity is the time-averaged deviation from sphericity over the system's characteristic timescales.

\section{Detection scheme}\label{app:detection}

\noindent
In what follows, we introduce an experimental scheme to measure the $z$ component of the angular momentum; the procedure can be analogously applied to the $x$ and $y$ components. The main purpose of the setup consists in the measurement of the position of the microspheres' centers of mass, which can be related to the angular momentum by virtue of both a non-uniform magnetic field used to trap the particle and the Barnett effect \cite{barnett}. Indeed, when considering the Hamiltonian associated to the center of mass of a single microsphere in the co-rotating frame, we have
\begin{equation}\label{comham}
\hat{H}_{c.o.m.}=\frac{\hat{p}^2}{2m}+\frac{|\chi_V| V}{2\mu_0}\lf\hat{\vec{B}}+\frac{\hat{\vec{L}}}{I\gamma}\ri^2-\hat{\vec{m}}\cdot\lf\hat{\vec{B}}+\frac{\hat{\vec{L}}}{I\gamma_p}\ri,
\end{equation}
where $\hat{\vec{B}}$ is the trapping magnetic field, $\mu_0$ the vacuum permittivity, $|\chi_V|$ the modulus of the magnetic susceptibility (negative for silica), $\gamma$ the gyromagnetic ratio specific for the considered material, $V$ the volume, $I$ the moment of inertia, $\hat{\vec{L}}$ the angular momentum, $\gamma_p$ the nuclear gyromagnetic ratio, $N$ the total number of nuclear spins, $\hat{\vec{S}}_i$ the respective spin operator for the $i$-th particle and
\begin{equation}
\hat{\vec{m}}=\gamma_p\sum_{i=1}^N\hat{\vec{S}}_i\,.
\end{equation}
We require the magnetic field to be non-uniform and dependent on the position of the center of mass:
\begin{equation}\label{B}
\hat{\vec{B}}=(0,0,\mathrm{G}_0 \hat{z}).
\end{equation}
where $\hat{z}$ is the position operator of the $z$-th coordinate. While such a magnetic field does not globally satisfy Maxwell's equations, physically realizable configurations -- such as quadrupole fields or anti-Helmholtz coils \cite{expdata}-- produce fields of the form $\vec{B}=(-G_0/2 \ \hat{x}, -G_0/2  \ \hat{y}, G_0 \  \hat{z})$. To ensure that the dominant field component is along z, the sphere's center of mass must remain predominantly along the z axis,\emph{ i.e.} $x , y \ll z$.\\
Furthermore, considering a magnetic field gradient of $G_0 = 10^6$ T$/$m, an angular velocity $\omega=10^7$ Hz and $\gamma = 8$ MHz/T, we have
\begin{align}\label{est}
    \expval{{\vec{\hat{B}}}^2} = G_0^2 \expval{\hat{z}^2} \approx 2.5\times10^3 \ \text{T}^2\,, 
    \qquad\frac{\expval{{\vec{\hat{L}}}^2}}{I^2 \gamma^2} = \frac{\omega^2}{\gamma^2}\approx 1.56 \ \text{T}^2\,,
\end{align}
therefore $\expval{{\vec{\hat{B}}}^2} \gg{\expval{{\vec{\hat{L}}}^2}}/{I^2 \gamma^2}$ and we can safely neglect the term quadratic in the angular momentum. In a similar fashion, one can check that the third term in Eq. \eqref{comham} is also negligible with respect to all the other contributions appearing in the Hamiltonian. Indeed, we see that, even in the worst-case scenario in which all the nuclear spins are pointing in the same direction (recall that $N=10^9$), one has
\begin{equation}
\expval{\hat{\vec{m}}\cdot\hat{\vec{B}}}\approx1.9\times10^{-26} \, \text{J},
\end{equation} 
while the lowest-order term which is relevant for our purposes is instead (with $|\chi_V|=1.13\times10^{-5}$)
\begin{equation}
\expval{\frac{|\chi_V| V}{2\mu_0}\hat{\vec{B}}\cdot\hat{\vec{L}}}\approx3.5\times10^{-11} \, \text{J}.
\end{equation}
Bearing this in mind, the Hamiltonian then becomes
\begin{equation}\label{comham2}
\hat{H}_{c.o.m.}=\frac{\hat{p}^2}{2m}+\frac{|\chi_V| V \mathrm{G}^2_0}{2\mu_0}\hat{z}^2+\frac{|\chi_V|V\mathrm{G}_0}{\gamma\mu_0I}\hat{
{z}}\hat{L}_z.
\end{equation}
Next, upon defining $\Omega=\sqrt{|\chi_V|\mathrm{G}^2_0/\rho\mu_0}$,
it is evident that one can identify in Eq. \eqref{comham2} the Hamiltonian of a harmonic oscillator, thus yielding 
\begin{equation}\label{comham3}
\hat{H}_{c.o.m.}=\hbar\Omega\hat{a}^\dagger\hat{a}+\hbar\lambda\hat{L}_z(\hat{a}^\dagger+\hat{a}),
\end{equation}
with $\lambda=|\chi_V|V\mathrm{G}_0/(\gamma I\mu_0\sqrt{2\hbar m\Omega})$.

At this point, it is possible to look at the time evolution of the position operator along the $z$ axis in the Heisenberg representation. Using the well-known commutation properties of the ladder operators appearing in \eqref{comham3}, we obtain
\begin{equation}\label{zop}
\hat{z}(t)=\hat{z}(0)+\sqrt{\frac{\hbar}{2m\Omega}}\frac{4\lambda\hat{L}_z(t)}{\Omega}\sin^2\lf\frac{\Omega t}{2}\ri.
\end{equation}
Therefore, the variance of the angular momentum can be obtained from the variance of the center of mass position along $z$ as follows 
\begin{align}\label{varang}
\Delta^2\hat{L}_z = \left(\frac{\gamma I \mathrm{G}_0}{2\sin^2\lf\frac{\Omega t}{2}\ri}\right)^2 \left( \Delta^2 \hat{z}(t) - \Delta^2 \hat{z}(0) \right),
\end{align}
where we assume that the positions are not time-correlated, \emph{i.e.} $\expval{\hat{z}(t) \hat{z}(0)} - \expval{\hat{z}(t)}\expval{\hat{z}(0)}=0$. Clearly, the above reasoning can be repeated for all the coordinate axes to obtain the expectation values of all the components of the angular momentum operator. 

This step is crucial if one wants to explicitly compute the sum uncertainty relation
\begin{equation}\label{sumun}
\sum_\alpha \lf \Delta^2 L_{A \alpha} + \Delta^2 L_{B \alpha} \ri \geq \hbar^2 \lf l_A + l_B \ri, \qquad \alpha=x,y,z,
\end{equation}
whose violation would yield information about the non-separability of the global state comprising $A$ and $B$. By resorting to the experimental values chosen for the microspheres (which can be found in the Letter), one can promptly verify that, in order to achieve the same sensitivity of the lower bound in \eqref{sumun} (namely, $10^{23} \hbar^2$), a Taylor expansion of Eq. \eqref{varang} provides
\begin{equation}\label{posun}
\Delta\hat{z}(t)\approx10^{-14}\,\mathrm{G}_0 t^2=10^{-12}\,\mathrm{G}_0.
\end{equation}
where in the second step we have set $t=10$ s.

In order to achieve an experimentally accessible spatial resolution, the magnetic field gradient must be considerably high. Along this line, it is worth stressing that, for sufficiently small regions of space, it is possible to achieve magnetic gradients as intense as $10^6$ T$/$m \cite{magneticgrad}, which in turn entails $\Delta\hat{z}\approx10^{-6}$ m. This value of the displacement error is definitely within the experimental reach; as a matter of fact, recent position measurements of levitated particles relying on optical and interferometric schemes {achieve  $1.7$ pm$/\sqrt{\text{Hz}}$} \cite{spatialmeasure}.

\section{Suppression of electric dipole moment}\label{app:dipole}
Assume that initially we know that the direction of the electric dipole moment of the microsphere is along the $x$ axis. We therefore rotate the sphere with an angular velocity $\omega_s$ around an orthogonal direction, say, $z$. The dipole moment vector is given by
\begin{align}\label{eq:dipvec}
    \vec{p}(t) = p \left( \cos \omega_s t \  \hat{x} +  \sin \omega_s t \  \hat{y}\right).
\end{align}
The time averaged dipole moment after a time $t_r$ is then
\begin{align}
    \expval{\vec{p}} = \frac{1}{t_r} \int_0^{t_r} \ \vec{p}(t) \text{d}t = \frac{p}{\omega_s t_r} \lf \sin \omega_s t_r \ \hat{x} + (1 - \cos  \omega_s t_r )\ \hat{y} \ri, 
\end{align}
which vanishes as $\omega_s t_r \gg 1$. For rotation periods much faster than the interaction time, the dipole magnitude is $\expval{p}:=\norm{\expval{\vec{p}}}\simeq p/\omega_s t_r$. Evaluating the potential energy due to the electric dipole-dipole we have 
\begin{align}
    V_{\text{dip-dip}} \simeq \frac{\expval{p} ^2 }{4 \pi \epsilon_0 r^3} = \frac{p^2 }{4 \pi \epsilon_0 \omega_s^2 t_r^2 r^3},
\end{align}
with $\epsilon_0$ the permittivity of free space. For a dipole moment of $p=100 \ e \ \mu$m \cite{afek}, a distance $r = 200 \mu$m, an angular velocity $\omega_s = 10^7 $ Hz and a time $t_r=1$ s, we have
\begin{align}
     V_{\text{dip-dip}} \approx 2.9 \times 10^{-39} \ \text{J}. 
\end{align}
Since the gravitational interaction we want to probe is $V_G\simeq10^{-38}$ J, the scheme described above is effective to make the electric dipole-dipole interaction subdominant with respect to gravity.\\
Now, suppose that it is not possible to prepare the microsphere's rotation axis perfectly orthogonal to the dipole's direction. The dipole vector has the following form:
\begin{align}
    \vec{p} (t) =  p \left( \cos \omega_s t \sin \delta \  \hat{x} +  \sin \omega_s t \sin \delta \  \hat{y} + \cos\delta \ \hat{z}\right).
\end{align}
Taking the time average, we find
\begin{align}\label{eq:ptilted}
    \norm {\expval{\vec{p}}}^2 = p^2 \lf \cos^2 \delta + \frac{\lf 2-2 \cos \omega_s t_r \ri \sin^2 \delta }{\omega_s^2 t_r^2} \ri.
\end{align}
Plugging in an angle $\delta = \pi/2 - 10^{-7}$ and the same parameters as above, gives an energy $V_{\text{dip-dip}} \simeq  10^{-39}$ J $< V_G$, \emph{i.e.} we can allow for deviations of the order of $\Delta \delta = 10^{-7}$ in order to keep the electric dipole-dipole energy subdominant with respect to gravity.

\section{Black-body radiation master equation}\label{app:blackbody}

\noindent
In order to derive a master equation that captures the effect of black-body radiation on our proposed 
experiment, we consider two bound quantum systems (atoms or molecules) that interact through the potential Eq. (4) of the main text while immersed in a quantized radiation field \cite{breuer}. Subsequently, we 
extend our analysis to a mesoscopic system, \emph{i.e.}, the SiO$_2$ microspheres. 

The starting point is the 
total Hamiltonian for the closed system, which includes the systems $A$ and $B$ as well as the environment, modeled 
as a thermal reservoir of bosonic modes
\begin{equation}
\hat{H}=\hat{H}_0+\hat{H}_E+\hat{H}_I,    
\end{equation}
where
\begin{eqnarray}\label{totham}
    \hat{H}_0 &=& \hat{h}_0\otimes \id + \id \otimes \hat{h}_0 \,, \; \;  \; \hat{h}_0 :=\frac{\hbar^2\hat{\vec{L}}^2}{2I}\,, \\[2mm]
    \hat{H}_E &=& \sum_\tk\hbar\omega_\tk \ha^\dagger_\tk \ha_\tk,\\[2mm]
    \hat{H}_I &=& \hbh_I^{AB}+\hbh_I^{ABE} \,,
\end{eqnarray}
where $\hat{\vec{L}}^2$ is a one-particle angular momentum operator, $I$ is the inertial moment of each sphere,
 $\omega_\tk=|\vk|c$ and the sum over $\tk$ also includes the polarizations of the electromagnetic field. Note that $\hat{H}_0$ acts on the subspace of the two particles $AB$.  If we consider the center of mass of spheres $A$ and $B$ to be located in $\vec{r}^A = \vec{0}$ and $\vec{r}^B=\vec{r_0}$ and denote their dipole moments as $\hvd^A$ and $\hvd^B$, respectively, we have
\begin{align}
   \hbh_I^{AB} &:= -\frac{\alpha \hbar}{2} \left(\hat{L}_{A+} \hat{L}_{B-} + \hat{L}_{A-} \hat{L}_{B+} -4\hat{L}_{Az} \hat{L}_{Bz}\right), \\[2mm] \hbh_I^{ABE} &=  -\lf\hvd^A \cdot{\hve \lf 0 \ri}+\hvd^B \cdot{\hve} \lf \vec{r}_{0} \ri \ri = -\lf\hvd^A \cdot{\hve }+\hvd^B \cdot{\hve} \ \text{e}^{i \vk \cdot \vec{r}_0} \ri  \nonumber \\[2mm] &\approx -\lf\hvd^A +\hvd^B  \ri \cdot \hve\,, \label{eq:HIABE}
\end{align}
where the operators labeled with $A(B)$ are understood to act only on the subspace of the particle $A(B)$. Note that the last approximation is only valid as long as $\vk \cdot \vec{r}_0 \ll 1$, meaning that the electric field wavelength is much larger than the separation between the spheres. In our case, we consider a center of mass separation of $\abs{\vec{r}_0}=2 \cdot 10^{-4}$ m and a characteristic wavelength of  $\lambda={hc}/{k_BT}\simeq{10^{-2}}/{T}$. Therefore, at $T=1$ K or below, the conditions for the approximation (\ref{eq:HIABE}) are fulfilled.  

On the other hand,  the electric field can be decomposed in the second quantization scheme as 
\begin{equation}\label{fieldexp}
\hve=i\sum_\tk\sqrt{\frac{2\pi\hbar\omega_\tk}{V \epsilon_0}}\,\vec{e}_\tk\lf\ha_\tk-\ha_\tk^\dagger\ri,    
\end{equation}
with $\vec{e}_\tk$ being the polarization vector, $V$ the volume of the box used to write the discretized expansion of the field and $\epsilon_0$ the permittivity of free space. 

Now, the Hamiltonian employed so far is the result of two underlying assumptions: i) the interaction between the thermal bath and each microsphere is treated as if each mesoscopic object is a big atom with appropriate angular momentum eigenstates and a suitably determined effective dipole strength;
ii) both microspheres interact with the same thermal bath. The first requirement is a simplification that allows us to write the dipole moment of each microsphere as 
\begin{equation}\label{dipmom}
\hvd^A=\hat{\vec{d}}\otimes\mathbb{1}, \qquad \hvd^B=\mathbb{1}\otimes \hat{\vec{d}}\,,  
\end{equation}
 where $\hat{\vec{d}}=q_{\mathrm{eff}}\hat{\vec{r}}$, $q_{\mathrm{eff}}$ is an effective charge to be estimated and $\hat{\vec{r}}$ the position operator, whilst the second requirement is reflected in the approximation (\ref{eq:HIABE}).
Next, moving to the interaction picture with respect to ${\hat H}_0$ and ${\hat H}_E$, we have
\begin{equation}\label{hamsplit}
\hbh_I(t)=\hbh_I^{AB}+\hbh_I^{ABE}(t).  
\end{equation}
 The equation for the reduced density matrix of the system $AB$ then becomes 
\begin{equation}\label{mesplit}
\frac{d}{dt}\hr_S(t)=-\frac{i}{\hbar}\left[\hbh_I^{AB},\hr_S(t)\right] - \frac{i}{\hbar}  \Tr_E\left[\hbh_I^{ABE} (t),\hr(t)\right],
\end{equation}
where $\hr_S:=\Tr_E \hr $ and $\hr $ is the density matrix of the full system $ABE$. From Eq. (\ref{mesplit}), we can already verify that the unitary part of the evolution only depends on the gravitational interaction between $A$ and $B$.
Bearing this in mind, let us insert the integral form of $\hr (t)$, given by
\begin{equation}
\hr(t)=\hr(0)-\frac{i}{\hbar}\int_0^tds\left[\hbh_I(s),\hr(s)\right],   
\end{equation}
into the second term of the r.h.s. in Eq. (\ref{mesplit}), thus obtaining
\begin{equation}\hspace{-12mm}
 -\frac{i}{\hbar}\Tr_E \left[\hbh_I^{ABE}(t),\hr(t)\right]=-\frac{1}{\hbar^2}\int_0^tds \Tr_E \left[\hbh_I^{ABE}(t),\left[\hbh_I(s),\hr(s)\right]\right]=-\frac{1}{\hbar^2}\int_0^tds \Tr_E \left[\hbh_I^{ABE}(t),\left[\hbh_I(s),\hr_S(t)\otimes\hr_E\right]\right],
\end{equation}
where we assume $\Tr_E \comm{\hbh_I^{ABE}(t)}{\hr (0)}=0$ and perform the Born approximation, \emph{i.e.}, $\hr(s)\approx\hr_S(s)\otimes\hr_E$ followed by the Markov approximation, according to which we replace $\hr (s)$ by $\hr (t)$ in order to have an equation local in time.

To achieve a Markovian master equation, we can substitute the variable $s$
with $t-s$ and push the extremal of the integral to infinity provided that the integrand goes to zero sufficiently fast \cite{breuer}. In light of this, Eq. (\ref{mesplit}) becomes
\begin{equation}\label{mid}
\frac{d}{dt}\hr_S(t)=-\frac{i}{\hbar}\left[\hbh_I^{AB},\hr_S(t)\right] -\frac{1}{\hbar^2}\int_0^\infty ds\,\mathrm{Tr}_E\left[\hbh_I^{ABE}(t),\left[\hbh_I(t-s),\hr_S(t)\otimes\hr_E\right]\right].    
\end{equation}
Assuming that $\hr_E$ is given by a thermal state, that is
\begin{equation}
\hr_E=\frac{\exp(-\beta\hbh_E)}{\mathrm{Tr}_E\exp(-\beta\hbh_E)}\,,  \qquad \beta=\frac{1}{k_BT}\,,   
\end{equation}
it is straightforward to check that
\begin{equation}
\left\langle\hve\right\rangle=\mathrm{Tr}_E\lf\hve\,\hr_E\ri=i\sum_\tk\sqrt{\frac{2\pi\hbar\omega_\tk}{V\epsilon_0}}\,\vec{e}_\tk\lf\left\langle\ha_\tk\right\rangle-\left\langle\ha^\dagger_\tk\right\rangle\ri=0,   
\end{equation}
thereby simplifying Eq. \eqref{mid} to
\begin{equation}\label{mid2}
\frac{d}{dt}\hr_S(t)=-\frac{i}{\hbar}\left[\hbh_I^{AB},\hr_S(t)\right] -\frac{1}{\hbar^2}\int_0^\infty ds\,\mathrm{Tr}_E\left[\hbh_I^{ABE}(t),\left[\hbh_I^{ABE}(t-s),\hr_S(t)\otimes\hr_E\right]\right],    
\end{equation}
since only quadratic terms in the environmental operators are non-vanishing.

In order to cast Eq. \eqref{mid2} in a Lindblad form, the recipe is to re-write the interaction Hamiltonian $\hbh_I^{ABE} (t)$ in terms of eigenoperators of the system $\hbh_0$ \cite{breuer}. First, note that the angular momentum states $\ket{l,m}$ are eigenkets of $\hat{h}_0$ with eigenvalues $E_{l} = {\hbar^2}l (l +1)/(2I)$. For simplicity, we are going to consider the same angular momentum for both spheres, meaning $l_A=l_B=l$. At this point, we define the one particle operator
\begin{align}\label{jump}
\hva(\Delta):=\sum_{\substack{m,m' \\ l,l'\,\mathrm{s.t.}\,E_{l'}-E_l=\frac{\hbar^2}{2I} \Delta}}|l,m\rangle\langle l,m|\hat{\vec{d}}|l',m'\rangle\langle l',m'|\,,    
\end{align} 
with $-l\leq m \leq l$, $-l'\leq m' \leq l'$ and where $\Delta$ is a fixed value. Note that $\Delta$ can only be an even integer.
Now, for both spheres we introduce the quantity 
\begin{align}
\hva^{AB} & := \hva \otimes \id + \id \otimes \hva  \\ &=  \sum_{\substack{m,m' \\[2mm] l,l'\,\mathrm{s.t.}\,E_{l'}-E_l=\frac{\hbar^2}{2I} \Delta}}  \bra{l,m}\hat{\vec{d}}\ket{l',m'} \Big [\ketbra{l,m}{l',m'}\otimes \id + \id \otimes \ketbra{l,m}{l',m'} \Big],
\end{align}
which is an eigenoperator of the system as
\begin{equation}
\left[\hat{h}_0,\hva(\Delta)\right]=-\frac{\hbar^2}{2I}\Delta\hva(\Delta), \qquad \left[\hat{h}_0,\hva^{\dagger}(\Delta)\right]=\frac{\hbar^2}{2I} \Delta\hva^{\dagger}(\Delta).
\end{equation}
One can then see that, when switching to the interaction picture, $\hva^{AB}$ only acquires a time-dependent phase
\begin{equation}\label{jumpint}
e^{\frac{i\hbh_0 t}{\hbar}}\hva^{AB}(\Delta)e^{-\frac{i\hbh_0 t}{\hbar}}=e^{-\frac{i t \hbar\Delta}{2I}}\hva^{AB}(\Delta), \qquad e^{\frac{i\hbh_0 t}{\hbar}}\hva^{AB\dagger}(\Delta)e^{-\frac{i\hbh_0 t}{\hbar}}=e^{\frac{i t \hbar \Delta}{2I}}\hva^{AB\dagger}(\Delta).
\end{equation}
Furthermore, by using the completeness relation for angular momentum states,
one can sum over $\Delta$ in Eq. (\ref{jump}) to obtain 
\begin{align}
    \sum_{\Delta\in 2\mathbb{Z}} \hva \lf \Delta \ri = \hat{\vec{d}}\,,
\end{align}
where $2\mathbb{Z}$ denotes the set of even integers. In the interaction picture, this entails
\begin{equation}\label{eq:dipole_intpic}
\hvd^{AB}(t)=\sum_{\Delta \in 2\mathbb{Z}}  e^{-\frac{i \hbar t \Delta}{2I}}\hva^{AB}(\Delta) =\sum_{\Delta\in 2\mathbb{Z}} e^{\frac{i \hbar t \Delta}{2I}}\hva^{AB\dagger}(\Delta),
\end{equation}
where we used $\hva^{AB\dagger}(\Delta)=\hva^{AB}(-\Delta)$.

We now have all the tools to 
convert Eq. \eqref{mid2} into a Lindblad-type master equation. By inserting the dipole moment (\ref{eq:dipole_intpic}) into Eq. \eqref{mid2} and working out the nested commutator, we arrive at
\begin{equation}\label{mid3}
\frac{d}{dt}\hr_S(t)=-\frac{i}{\hbar}\left[\hbh_I^{AB},\hr_S(t)\right]+ \sum_{\substack{\Delta,\Delta' \\ p,q=1,2,3 }}e^{\frac{i\hbar (\Delta'-\Delta)t}{2I}}\Gamma_{pq}(\Delta)\left[\hba^{AB}_q(\Delta)\hr_S(t)\hba^{AB\dagger}_p(\Delta')-\hba^{AB\dagger}_p(\Delta')\hba^{AB}_q(\Delta)\hr_S(t)\right]+\mathrm{h.c.},
\end{equation}
where we implicitly assume that the sum is carried over the even integers, $\hba^{AB}_p$ denotes the $p$-th component of $\hva^{AB}$ in some orthonormal basis $\{\vec{v}_1, \vec{v}_2, \vec{v}_3 \}$ and
\begin{equation}
\Gamma_{pq}(\Delta)=\int_0^\infty ds e^{\frac{i\hbar s\Delta}{2I}}\left\langle\hbe_p(t)\hbe_q(t-s)\right\rangle\,.
\end{equation}
Noting that on a thermal state $\left\langle\hbe_p(t)\hbe_q(t-s)\right\rangle\ = \left\langle\hbe_p(s)\hbe_q(0)\right\rangle\ $, we can perform the rotating wave approximation in Eq. \eqref{mid3}, where we only consider terms for which $\Delta'=\Delta$. This holds true because the phase factors in Eq. \eqref{mid3} oscillate very rapidly in time intervals where the operator $\hr_S(t)$ changes appreciably. Bearing this in mind, we write
\begin{equation}\label{mid4}
\frac{d}{dt}\hr_S(t)=-\frac{i}{\hbar}\left[\hbh_I^{AB},\hr_S(t)\right]+ \sum_{\substack{\Delta \\ p,q=1,2,3}}\Gamma_{pq}(\Delta)\left[\hba^{AB}_q(\Delta)\hr_S(t)\hba^{AB\dagger}_p(\Delta)-\hba^{AB\dagger}_p(\Delta)\hba^{AB}_q(\Delta)\hr_S(t)\right]+\mathrm{h.c.}.
\end{equation}
At this stage, there is no substantial difference with respect to the standard treatment of open quantum systems when it comes to evaluate the factor $\Gamma_{pq}(\Delta)$. In particular, as we chose a thermal reservoir, the correlation functions of the electric field can be easily computed \cite{breuer,fogedby} and, if we neglect the contributions associated to the renormalization of the system Hamiltonian (namely, Lamb and Stark shift terms), we arrive at the expression 
\begin{align}\label{ap:ME}
\frac{d}{dt}\hr_S(t) = &-\frac{i}{\hbar}\left[\hbh_I^{AB},\hr_S(t)\right] \nonumber \\[2mm] &+ \sum_{\substack{\Delta>0}}\frac{\Delta ^3 \hbar^2}{6 c^3 I^3 \epsilon_0}\lf1+N(\Delta)\ri\left[\hva^{AB}(\Delta)\cdot\hr_S(t)\hva^{AB\dagger}(\Delta)-\frac{1}{2}\left\{\hva^{AB\dagger}(\Delta)\cdot\hva^{AB}(\Delta),\hr_S(t)\right\}\right] \nonumber \\[2mm] &+\sum_{\substack{\Delta>0 }}\frac{\Delta ^3 \hbar^2}{6 c^3 I^3 \epsilon_0} N(\Delta)\left[\hva^{AB\dagger}(\Delta)\cdot\hr_S(t)\hva^{AB}(\Delta)-\frac{1}{2}\left\{\hva^{AB}(\Delta)\cdot\hva^{AB\dagger}(\Delta),\hr_S(t)\right\}\right], 
\end{align}
where 
\begin{equation}\label{occ}
N(\Delta)=\frac{1}{\exp \lf \frac{\beta \hbar^2 \Delta}{2I} \ri-1 }\,,
\end{equation}
is the average number of photons  with energy ${\hbar^2 \Delta}/{2I}$ as given by the Planck distribution. Although this is the final expression for our master equation, there is still one last task to complete. As a matter of fact, we have not yet provided the matrix element of the dipole moment in the angular momentum representation $\langle l,m|\hat{\vec{d}}|l',m'\rangle$, which turns out to be crucial to compute the jump operators \eqref{jump}. In spherical coordinates, we have
\begin{equation}\label{rexp}
\langle l,m|\hat{\vec{d}}|l',m'\rangle=\qe R\lf\langle l,m|\sin\hth\cos\hvp|l',m'\rangle\vec{e}_x+\langle l,m|\sin\hth\sin\hvp|l',m'\rangle\vec{e}_y+\langle l,m|\cos\hth|l',m'\rangle\vec{e}_z\ri,
\end{equation}
where $R$ is the radius of both spheres $A$ and $B$. In general, when computing the expectation value of a function of the spherical angles $f(\hth,\hvp)$ on angular momentum eigenstates, it is convenient to introduce the direction eigenkets $|\vn\rangle$, which are eigenvectors of the angles $\hth$ and $\hvp$ with eigenvalues $\theta$ and $\varphi$, respectively. Furthermore, these states fulfill the completeness relation \cite{sakurai}
\begin{equation}
\int d\Omega|\vn\rangle\langle\vn|=\int_0^{2\pi} d\varphi\int_0^\pi\sin\theta d\theta|\vn\rangle\langle\vn|=\mathbb{1}\,.
\end{equation}
In light of this, it is possible to consider the expectation value of a generic $f(\hth,\hvp)$ as
\begin{equation}
\langle l,m|f(\hth,\hvp)|l',m'\rangle=\int d\Omega\int d\Omega'\langle l,m|\vn\rangle\langle\vn|f(\hth,\hvp)|\vn'\rangle\langle\vn'|l',m'\rangle=\int d\Omega f(\theta,\varphi)\langle l,m|\vn\rangle\langle\vn|l',m'\rangle\,.
\end{equation}
In the previous expression, one can recognize the emergence of the spherical harmonics \cite{sakurai}, since $\langle\vn|l',m'\rangle=Y^{m'}_{l'}(\theta,\varphi)$. Thus, the whole computation amounts to an evaluation of an integral over the solid angle of products of spherical harmonics. Indeed, the functions of the spherical angles in Eq. \eqref{rexp} can be written as linear superpositions of spherical harmonics with $l=1$ as 
\begin{equation}
\sin\theta\cos\varphi=\sqrt{\frac{2\pi}{3}}\left[Y^{-1}_1(\theta,\varphi)-Y^{1}_1(\theta,\varphi)\right], \quad \sin\theta\sin\varphi=i\sqrt{\frac{2\pi}{3}}\left[Y^{-1}_1(\theta,\varphi)+Y^{1}_1(\theta,\varphi)\right], \quad \cos\theta=\sqrt{\frac{4\pi}{3}}Y^{0}_1(\theta,\varphi).
\end{equation}
With this knowledge, we can now solve the integrals by virtue of the formula \cite{grad}
\begin{equation}
\int d\Omega Y^{m_1}_{l_1}(\theta,\varphi)Y^{m_2}_{l_2}(\theta,\varphi)Y^{m_3}_{l_3}(\theta,\varphi)=\sqrt{\frac{(2l_1+1)(2l_2+1)(2l_3+1)}{4\pi}}\begin{pmatrix} l_1 & l_2 & l_3 \\ 0 & 0 & 0  \end{pmatrix}\begin{pmatrix} l_1 & l_2 & l_3 \\ m_1 & m_2 & m_3  \end{pmatrix},
\end{equation}
where we identify the Wigner 3-j symbol
\begin{equation}
\begin{pmatrix} l_1 & l_2 & l_3 \\ m_1 & m_2 & m_3  \end{pmatrix}=\frac{(-1)^{l_1-l_2-m_3}}{\sqrt{2l_3+1}}\,C^{l_3-m_3}_{l_1m_1l_2m_2}\,,
\end{equation}
with $C$ being the Clebsch-Gordan coefficients. \\
Finally, using the property $(Y^m_l(\theta,\varphi))^*=(-1)^mY^{-m}_l(\theta,\varphi)$, we can cast the transition amplitude of the dipole moment $\hat{\vec{d}}$ in the following form:
\begin{align}\label{eq:dipole_trans}
 &\bra{l, m} \hat{\vec{d}}\ket{l', m'} = \nonumber \\[2mm] & (-1)^m\qe R\sqrt{(2l+1)(2l'+1)}\begin{pmatrix} 1 & l & l' \\ 0 & 0 & 0  \end{pmatrix}\Bigg[\begin{pmatrix} 1 & l & l' \\ -1 & -m & m'  \end{pmatrix}\frac{\vec{e}_x+i\vec{e}_y}{\sqrt{2}}+i\begin{pmatrix} 1 & l & l' \\ 1 & -m & m'  \end{pmatrix}\frac{i\vec{e}_x+\vec{e}_y}{\sqrt{2}} \nonumber\\[2mm] &+\begin{pmatrix} 1 & l & l' \\ 0 & -m & m'  \end{pmatrix}\vec{e}_z\Bigg]. 
\end{align}
It is worth stressing that the assumption of an atom-like structure for the microspheres has led to the employment of a simple form for the dipole moments. However, to account for a more realistic scenario, the effective charge $\qe$ has to be properly estimated and must include all the effects that pertain to a mesoscopic body.

The Wigner 3-j symbols carry the following selection rules
\begin{enumerate}
    \item $\begin{pmatrix} 1 & l & l' \\ 0 & 0 & 0  \end{pmatrix} \rightarrow \abs{l-1} \leq l' \leq l+1 $,

    \item $\begin{pmatrix} 1 & l & l' \\ -1 & -m & m'  \end{pmatrix} \rightarrow m'=m+1$,

    \item $\begin{pmatrix} 1 & l & l' \\ 1 & -m & m'  \end{pmatrix} \rightarrow m'=m-1$,

    \item $\begin{pmatrix} 1 & l & l' \\ 0 & -m & m'  \end{pmatrix} \rightarrow m'=m$.
\end{enumerate}
Since Eq. (\ref{ap:ME}) only requires the jump operators (\ref{jump}) for which $\Delta>0$, we have
\begin{align}\label{ap:vecA}
\hva(\Delta>0)=d_{\text{eff}} \sum_{\substack{m,m' \\ l,l'\,\mathrm{s.t.}\, E_{l'}-E_{l}=\frac{\hbar^2}{2I} \Delta}} & M_{l,l',m}  \begin{pmatrix} 1 & l & l' \\ 0 & 0 & 0  \end{pmatrix} \Bigg[\begin{pmatrix} 1 & l & l' \\ -1 & -m & m'  \end{pmatrix}\vec{v}_1+i\begin{pmatrix} 1 & l & l' \\ 1 & -m & m'  \end{pmatrix}\vec{v}_2 \nonumber\\[2mm] &+\begin{pmatrix} 1 & l & l' \\ 0 & -m & m'  \end{pmatrix}\vec{v}_3\Bigg] \ketbra{l,m}{l',m'},
\end{align} 
where $d_{\text{eff}}:=\qe R$, $M_{l,l',m}:= (-1)^m \sqrt{(2l+1)(2l'+1)}$ and we define the rotated basis $\{\vec{v}_1 = {(\vec{e}_x+i\vec{e}_y})/{\sqrt{2}}, \vec{v}_2={ (i\vec{e}_x+\vec{e}_y)}/{\sqrt{2}}, \vec{v}_3=\vec{e}_z \}$. The selection rule number (1), along with the fact that $E_{l'}-E_l>0$ and that $l,l'$ non-negative are integers imply that the only non-vanishing contributions to the sum are those with $l'=l+1$. Since $\Delta$ is a fixed value, only the term with $l = \Delta/2 - 1$ contributes to the sum. Then, including the selection rules for $m$ we have 
\begin{align} 
\hat{A}_1 \lf \Delta \ri &= \sum_{m=-l }^{l } M_{l,l+1,m} \begin{pmatrix} 1 & l & l +1 \\ 0 & 0 & 0  \end{pmatrix} \begin{pmatrix} 1 & l  & l +1 \\ -1 & -m & m+1  \end{pmatrix} \ketbra{l ,m}{l +1,m+1} \label{ap:A1}\\[2mm] \hat{A}_2 \lf \Delta \ri &= \sum_{m=-l }^{l } i M_{l ,l +1,m} \begin{pmatrix} 1 & l  & l +1 \\ 0 & 0 & 0  \end{pmatrix} \begin{pmatrix} 1 & l  & l +1 \\ 1 & -m & m-1  \end{pmatrix} \ketbra{l ,m}{l +1,m-1} \label{ap:A2} \\[2mm] \hat{A}_3 \lf \Delta \ri &= \sum_{m=-l }^{l } M_{l ,l +1,m} \begin{pmatrix} 1 & l  & l +1 \\ 0 & 0 & 0  \end{pmatrix} \begin{pmatrix} 1 & l  & l +1 \\ 0 & -m & m  \end{pmatrix} \ketbra{l ,m}{l +1,m}  \label{ap:A3}
\end{align}
where $\hat{A}_k$ is the $k$-th component of vector (\ref{ap:vecA}) and for simplicity we have omitted that $l =l (\Delta)$.\\
Finally, since there is a one-to-one correspondence between $l$ and $\Delta$, we can change the summation index in Eq. (\ref{ap:ME}) to write
\begin{align}\label{ap:MEfinal}
\frac{d}{dt}\hr_S(t) = &-\frac{i}{\hbar}\left[\hbh_I^{AB},\hr_S(t)\right] \nonumber \\[2mm] &+ \sum_{\substack{l\geq 0}}\frac{\Delta_l ^3 \hbar^2}{6 c^3 I^3 \epsilon_0}\lf1+N(\Delta_l)\ri\left[\hva^{AB}(\Delta_l)\cdot\hr_S(t)\hva^{AB\dagger}(\Delta_l)-\frac{1}{2}\left\{\hva^{AB\dagger}(\Delta_l)\cdot\hva^{AB}(\Delta_l),\hr_S(t)\right\}\right] \nonumber \\[2mm] &+\sum_{\substack{l\geq 0 }}\frac{\Delta_l ^3 \hbar^2}{6 c^3 I^3 \epsilon_0} N(\Delta_l)\left[\hva^{AB\dagger}(\Delta_l)\cdot\hr_S(t)\hva^{AB}(\Delta_l)-\frac{1}{2}\left\{\hva^{AB}(\Delta_l)\cdot\hva^{AB\dagger}(\Delta_l),\hr_S(t)\right\}\right], 
\end{align}
where $\Delta_l := 2(l+1)$ and $\hva^{AB} = \sum_i \hat{A}_i \vec{v}_i \otimes \id + \id \otimes  \sum_i \hat{A}_i \vec{v}_i $ as given by Eqs. (\ref{ap:A1}-\ref{ap:A3}).

\clearpage

\end{widetext}

\end{document}